\RequirePackage{fix-cm}
\documentclass[twocolumn,natbib]{arxivtwocolumn}
\smartqed  
\usepackage{graphicx}
\usepackage{mathptmx}      

\usepackage[utf8]{inputenc}
\usepackage{amsmath}
\usepackage{amsfonts}
\usepackage{amssymb}
\usepackage{answers}
\usepackage{subfig}
\usepackage{stmaryrd}
\usepackage{xparse}
\usepackage{letltxmacro}
\usepackage{algpseudocode}
\usepackage{algorithm}
\usepackage{numprint}
\usepackage{import}
\usepackage[inline]{enumitem}
\usepackage{xfrac}
\usepackage{color}
\usepackage[english]{babel}

\Newassociation{proofprop}{demprop}{ann}

\Newassociation{prooflem}{demlem}{ann}

\Newassociation{proofcoro}{demcoro}{ann}

\newcommand{\E}[1]{\operatorname{E}\left[ #1 \right]}
\newcommand{\var}[1]{\operatorname{var}\left[ #1 \right]}
\newcommand{\proba}[1]{\operatorname{P}\left[ #1 \right]}
\usepackage{bbm}
\newcommand{\one}{\mathbbm{1}}
\newcommand{\U}{\ensuremath{\mathbf{U}}}

\renewcommand{\u}{\ensuremath{\mathbf{u}}}
\newcommand{\R}{\ensuremath{\mathbb{R}}}
\newcommand{\N}{\ensuremath{\mathbb{N}}}
\newcommand{\I}[2]{\displaystyle\int_{#1}^{#2}}
\newcommand{\dx}{\ensuremath{\mathrm{d}x}}
\renewcommand{\d}{\ensuremath{\mathrm{d}}}
\newcommand{\loi}{\overset{\mathcal{L}}{\sim}}
\renewcommand{\=}{\overset{\text{def}}{=}}
\newcommand{\mmck}{\widehat{m}_{MC}}

\newcommand{\m}{\widehat{m}}

\NewDocumentCommand{\Sum}{O{i = 1}O{N}}{\sum \limits_{#1}^{#2}}
\renewcommand{\l}{\left(}
\renewcommand{\r}{\right)}
\newcommand{\lb}{\left[}
\newcommand{\rb}{\right]}
\newcommand{\burnin}{\textit{burn-in }}
\newcommand{\argmin}{\operatorname{arg min}}
\newcommand{\mis}{\widehat{m}_{IS}}
\newcommand{\zk}{\widehat{Z}}
\newcommand{\z}{\widehat{Z}}
\newcommand{\Z}{\widehat{Z}}
\newcommand{\alp}{\widehat{\alpha}}
\LetLtxMacro{\oldsubsection}{\subsection}
\renewcommand{\subsection}[1]{\oldsubsection{#1} \label{ss:#1}}
\LetLtxMacro{\oldsection}{\section}
\renewcommand{\section}[1]{\oldsection{#1} \label{s:#1}}
\newcommand{\appenproof}{\oldsection*{Appendix}}
\newcommand{\ie}{\textit{i.e. }}
\newcommand{\eg}{\textit{e.g. }}
\newcommand{\cdf}{\textit{cdf}}
\newcommand{\pdf}{\textit{pdf }}
\newcommand{\iid}{\textit{iid }}

\journalname{Statistics and Computing}

\begin{document}

\title{Point Process-based Monte Carlo estimation}
\author{Cl\'ement Walter}
\institute{C. Walter \at
              CEA, DAM, DIF, F-91297 Arpajon, France \\
              Tel.: +331-69-264000\\
              \email{clement.walter@cea.fr}
           \and
           C. Walter \at
              Laboratoire de Probabilités et Modèles Aléatoires \\ Université Paris Diderot, Paris, France
}

\date{Laboratoire de Probabilités et Modèles Aléatoires \\ Université Paris Diderot, Paris, France \\
      CEA, DAM, DIF, F-91297 Arpajon, France \\
              Tel.: +331-69-264000\\
              \email{clement.walter@cea.fr} \\
}


\maketitle

\begin{abstract}
This paper addresses the issue of estimating the expectation of a real-valued random variable of the form $X = g(\U)$ where $g$ is a deterministic function and $\U$ can be a random finite- or infinite-dimensional vector. Using recent results on rare event simulation, we propose a unified framework for dealing with both probability and mean estimation for such random variables, \ie linking algorithms such as Tootsie Pop Algorithm (TPA) or Last Particle Algorithm with nested sampling. Especially, it extends nested sampling as follows: first the random variable $X$ does not need to be bounded any more: it gives the principle of an ideal estimator with an infinite number of terms that is unbiased and always better than a classical Monte Carlo estimator -- in particular it has a finite variance as soon as there exists $k \in \R > 1$ such that $\E{X^k} < \infty$. Moreover we address the issue of nested sampling termination and show that a random truncation of the sum can preserve unbiasedness while increasing the variance only by a factor up to 2 compared to the ideal case. We also build an unbiased estimator with fixed computational budget which supports a Central Limit Theorem and discuss parallel implementation of nested sampling, which can dramatically reduce its computational cost. Finally we extensively study the case where $X$ is heavy-tailed.

\end{abstract}

\Opensolutionfile{ann}[demo]

\section{Introduction}
Nested sampling was introduced in the Bayesian framework by \citet{skilling2006nested} as a method for ``estimating directly how the likelihood function relates to prior mass". Formally, it builds an approximation for the evidence:
\[
Z = \I{\Theta}{} L(\theta) \pi (\theta) \d \theta,
\]
where $\pi$ is the prior distribution, $L$ the likelihood, and $\Theta \subset \R^d$. It is somehow a quadrature formula but in the $[0,1]$ interval rather than in the original multidimensional space $\Theta$:
\begin{align*}
Z &= \I{0}{1} Q(P) \d P,
\end{align*}
where $Q$ is the quantile function which is the generalised inverse of:
\[
P(\lambda) = \I{L(\theta)>\lambda}{} \pi(\theta) \d \theta .
\]
Hence the name \emph{nested sampling} because the initial input space is divided into nested subsets $\{\theta \in \Theta \mid L(\theta) > \lambda \}$. Convergence of the approximation error toward a Gaussian distribution has been proved \citep{chopin2010properties} when assuming that $Q$ is twice continuously differentiable with its two first derivatives bounded over $[\varepsilon, 1]$ for some $\varepsilon > 0$.

On the other hand estimating a quantity such as $P(\lambda)$ for a given $\lambda$ is a typical problem arising in rare event probability estimation. In this context, $L$ (often denoted by $g$) represents a complex computer code (not necessarily positive valued nor continuous nor bounded), $\theta$ is a vector of parameters, and $F_\lambda = \{ \theta \in \Theta \mid L(\theta) > \lambda \}$ is the so-called failure domain. The idea of writing $F_\lambda$ as a finite intersection of nested subsets $F_{\lambda_0} \supset \cdots \supset F_{\lambda_n},\; -\infty = \lambda_0 < \cdots < \lambda_n = \lambda$ goes back to \citet{kahn1951estimation} and is now referred to as Multilevel Splitting \citep{garvels2000splitting,cerou2007adaptive} or Subset Simulation \citep{au2001estimation}. Statistical properties and convergence results have been derived by interpreting the Splitting algorithm in terms of an Interacting Particles System \citep{cerou2009rare,cerou2012sequential}. Furthermore a particular implementation, sometimes called the \emph{Last Particle Algorithm} (LPA), has gained a lot of attention and \citet{huber2011using,huber2014random}, \citet{guyader2011simulation} and \citet{simonnet2014combinatorial} have independently proved its link with a Poisson process. This algorithm is indeed somehow the one proposed by \citet[Section 6]{skilling2006nested} but the connection between nested sampling and rare event simulation remains unclear (see \citet{guyader2011simulation} and the discussion following \citet{huber2011using} in \citet{bernardo2011bayesian}).

The goal of this paper is to fill this gap by introducing a common framework for these methodologies. The core tool is that any continuous real-valued random variable can be linked with a Poisson process with parameter 1. Then a family of estimators can be defined using several realisations of such processes instead of \iid samples. While it only recasts results for extreme probability estimation in a very general setting -- \ie the random variable of interest writes as $X = g(\U) \in \R$ where $g$ is a deterministic function and $\U$ can be a random finite- or infinite-dimensional vector -- it extends nested sampling to the estimation of the mean of any real-valued random variables (bounded or not) and brings new theoretical results:
\begin{enumerate*}[label=\arabic*)]
\item the ideal estimator with an infinite number of terms (non truncated nested sampling) is unbiased;
\item the ideal nested sampling estimator is always better than the classical Monte Carlo estimator in term of variance; and 
\item it has a finite variance as soon as a moment of order $k \in (1, \infty)$ exists.
\end{enumerate*}

Moreover we address the issue of the nested sampling termination \citep[see][Section 7]{skilling2006nested}. Using results on Multilevel Monte Carlo \citep{giles2008multilevel, mcleish2011ageneral,rhee2013unbiased}, we show that one can get an unbiased estimator with a random but a.s. finite number of terms whose variance is only twice the one of the ideal estimator. We also build an unbiased estimator with a fixed computational budget which supports a Central Limit Theorem. We further discuss parallel implementation of nested sampling and these new estimators as this can can dramatically reduce its computational cost.

All these theoretical results are derived assuming that it is possible to generate samples according to conditional laws when it is required. This is indeed a tough requirement but this problem is well identified and not particular to these randomised estimators \citep[see][]{roberts2011comment:TPA}; especially \citet{skilling2006nested,huber2011using,guyader2011simulation} already acknowledge it and make use of Markov Chain Monte Carlo sampling. While a lot of ongoing work on nested sampling focus on improving these conditional simulations \citep[\eg][]{brewer2011diffusive}, in the present article we focus on theoretical statistical properties and suggest a possible solution to the issue of choosing a \emph{bad} stopping criterion.
Hence, it is out of the scope of the present work to benchmark nested sampling against other tailor-made methods such as Importance Sampling (see for example \citep{robert2004monte} or \citep{glynn1989importance}) on a list of specific cases.

The outline of this paper is as follows: Section \ref{s:Ideal estimator} presents the common framework for rare event simulation and nested sampling and derives a new \emph{ideal} (not practically implementable) estimator of $m = \E{X} = \E{g(\U)}$. It is closely related to nested sampling with an infinite number of terms and is compared to the usual Monte Carlo estimator. Section \ref{s:Randomised unbiased estimator} proposes two possible estimators based on the \emph{ideal} one. Section \ref{s:Application to heavy-tailed random variables} studies the specific case where $X = g(\U)$ is heavy-tailed and Section \ref{s:Example} gives information on practical implementation and numerical results. Finally an Appendix gathers all the proofs.

\section{Ideal estimator} \label{section_moments_def}

From now on we consider a real-valued random variable $X$, which can be for instance the output of a mapping $X = g(\U)$, as discussed in the Introduction.

Furthermore for a real-valued random variable $X$, one can write $X = X_+ - X_-$ with $X_+$ and $X_-$ non-negative random variables. Then, $\E{X} = \E{X_+} - \E{X_-}$. Thus in the sequel and without loss of generality we assume that $X$ is a non-negative random variable with law $\mu^X$. We also assume that $X$ has a continuous \textit{cdf} $F$ and we write $p_x$ instead of $\proba{X > x} = 1 - F(x)$, for any $x \in \R^+$.

\subsection{Extreme event simulation} \label{ss:exteme event simulation}

In this section we recast common results from \citep{huber2011using,guyader2011simulation,simonnet2014combinatorial} in a general framework.

\begin{definition}[Increasing random walk] \label{def:increasing rw}
Let $X_0 = 0$ and define recursively the Markov sequence $(X_n)_n$ such that
\begin{equation*}
\forall n \in \N : \proba{X_{n+1} \in A \mid X_0, \cdots, X_n} = \dfrac{\mu^X(A \cap (X_n, +\infty) ) }{\mu^X( (X_n, +\infty) )}.
\end{equation*}
\end{definition}
In other words $(X_n)_n$ is a strictly increasing sequence where each element is generated conditionally greater than the previous one. Considering the sequence $(T_n)_{n\geq1}$ such that $T_n = -\log \l \proba{X > X_n} \r $, it can be shown that $(T_n)_{n\geq1}$ is distributed as the arrival times of a Poisson Process with parameter 1. Thus, the counting random variable of the number of events before $x$: $M_x = \operatorname{card} \{ n \geq 1 \mid X_n \leq x \}$ follows a Poisson law with parameter $t_x = -\log p_x$.

This result leads to the construction of a new estimator for the probability of exceeding a threshold $x$. Indeed Lehmann-Scheffé theorem states that the minimum-variance unbiased estimator (MVUE) for $p_x = e^{-{t_x}}$ is
\begin{equation} \label{eq_proba_estimator}
\widehat{p_x} = \l 1 - \dfrac{1}{N} \r ^{M}
\end{equation}
with $M = \sum_{i=1}^N M_x^i$ the sum of $N$ \iid realisations of $M_x$. Here we find back the LPA estimator, which means that LPA is only one possible practical implementation of this estimator; especially \citet{walter2015moving} shows that LPA generates a marked Poisson Process with parameter $N$. In any case, the statistical properties of $\widehat{p_x}$ are then well known:
\begin{proposition}[Statistical properties of $\widehat{p_x}$] \label{propo:stat proba estimator}
\begin{align*}
\E{\widehat{p_x}} &= p_x \\
\var{\widehat{p_x}} &= p_x^2 \l p_x^{-1/N} - 1 \r 
\end{align*}
\end{proposition}
This estimator exhibits a logarithmic efficiency and asymptotically achieves the Cramer-Rao bound $-p_x^2 \log p_x / N$. Comparing to classical Monte Carlo, it \emph{replaces} the factor $1/p_x$ in the variance by $\log 1/p_x$ when $p_x \ll 1$ and $N \gg 1$:
\vskip6pt
\begin{tabular}{l|c|c}
& classical Monte Carlo & Poisson Process \\
\hline
\rule[-15pt]{0pt}{35pt}Variance & $\dfrac{p_x(1-p_x)}{N}$ & $p_x^2 \l p_x^{-1/N} - 1 \r$ \\
Approx. & $\dfrac{p_x^2}{N} \dfrac{1}{p_x}$ & $\dfrac{p_x^2}{N} \log \dfrac{1}{p_x}$
\end{tabular}
\vskip6pt
\begin{remark} \label{rem:suboptimal p estimator}
The MVUE of $t_x = - \log p_x$ is $M/N$. From this relation one could consider the suboptimal estimator for $p_x$:
\begin{equation}
\widetilde{p_x} = e^{-\frac{M}{N}} = \l e^{-\frac{1}{N}} \r^M .
\end{equation}From the moment-generating function of $M$ we get the mean and variance of $\tilde{p}_x$:
\begin{align*}
\E{\widetilde{p_x}} &= p_x^{N ( 1 - e^{-1/N} )} = p_x + \dfrac{- p_x \log p_x}{2 N} + o \l \dfrac{1}{N} \r \\
\var{\widehat{p_x}} &= p_x^{N ( 1 - e^{-2/N} )} - p_x^{2 N ( 1 - e^{-1/N} )} \\
&= \dfrac{- p_x^2 \log p_x}{N} + \dfrac{p_x^2 \log p_x}{N^2} \l \log p_x + 1 \r + o\l \dfrac{1}{N^2} \r .
\end{align*}
Hence this suboptimal estimator has a positive bias of order $1/N$. The variances $\var{\widetilde{p_x}}$ and $\var{\widehat{p_x}}$ differ only from order $1/N^2$ and $\var{\widehat{p_x}} < \var{\widetilde{p_x}}$ as soon as $p_x < e^{-1}$.
\end{remark}

\subsection{Definition of the moment estimator}
Noticing that for a non-negative real-valued random variable with mean $m = \E{X} = \E{g(\U)}$ one has:
\begin{equation}
m = \I{0}{\infty} p_x \dx ,
\end{equation}
the idea is to use the optimal estimator of $p_x$ (Eq. \eqref{eq_proba_estimator}) to build an estimator for $m$.

From now on we will assume that $N \geq 2$ point processes have been simulated and denote by $(M_x)_x$ the counting random variables associated with the marked Poisson Process: $\forall x > 0, M_x \sim  \mathcal{P}(-N \log p_x)$. The sequence $(X_n)_{n\geq1}$ is the cumulated one, \ie the combination of the states of the $N$ Markov Chains sorted in increasing order; then the associated $(T_n)_{n \geq 1}$ are the times of the marked Poisson Process with parameter $N$. We set $X_0 = 0$ and then consider the following estimator:
\begin{align} \label{eq:mk def}
\m &= \I{0}{\infty} \l 1 - \dfrac{1}{N} \r ^{M_x} \dx \notag \\
  &= \sum \limits_{i=0}^\infty \l X_{i+1} - X_i \r  \l  1 - \dfrac{1}{N} \r ^i .
\end{align}
The second equality comes from the fact that $x \mapsto M_x$ is constant equal to $i$ on each interval $[X_i, X_{i+1})$: there are $0$ event before $X_1$, then $1$ event before $X_2$, precisely at $X_1$, etc.

While the first form is easier to analyse because the law of $(M_x)_x$ is well determined, the second one paves the way for the practical implementation (see Section \ref{s:Randomised unbiased estimator}) and clarifies the link with Nested Sampling:
\begin{align} \label{eq:mk link nested}
\m
&= \sum \limits_{i=1}^\infty X_i \left[ \l  1 - \dfrac{1}{N} \r^{i-1} - \l  1 - \dfrac{1}{N} \r^{i} \right] .
\end{align}
This estimator is the limit of the nested sampling estimator with a deterministic scheme \citep{skilling2006nested}:
\begin{equation} \label{eq:original infinite nested sampling}
\widetilde{m} = \Sum[i=1][\infty] X_i \l e^\frac{1-i}{N} - e^\frac{-i}{N} \r
\end{equation}
with slightly modified weights: $(1-1/N)$ instead of $e^{-1/N}$. This is a direct consequence of the fact that an optimal unbiased estimator for $e^{-t_x}$ is not $e^{-\hat{t_x}}$ (see Section \ref{ss:Extreme event simulation} Remark \ref{rem:suboptimal p estimator}).

\begin{proposition}[Statistical properties of $\m$] \label{propo_stat_estimateur}
\begin{align}
\E{\m} &= m \\
\var{\m} &= 2 \int_0^\infty \int_0^x p_x p_{x'}^{1-1/N}\dx' \dx - m^2 \label{eq:variance m}
\end{align}
\begin{proofprop}
one has:
\[
\E{\m} =  \int_0^\infty \E{  \left( 1 - \dfrac{1}{N} \right)^{M_x} } \dx = \int_0^\infty p_x \dx .
\]
For the variance, one uses the fact that, for $x>x'$, $M_x - M_{x'}$ and $M_{x'}$ are independent to expand $\E{\m^2}$:
\begin{multline*}
\E{\m^2} = 2 \int_0^\infty \int_0^x \E{\left(1-\dfrac{1}{N} \right)^{M_x + M_{x'}}} \dx' \dx \\
= \int_0^\infty \int_0^x \E{\left(1-\dfrac{1}{N} \right)^{M_x - M_{x'}} \left(1-\dfrac{1}{N} \right)^{2 M_{x'}}} \dx' \dx .
\end{multline*}
Furthermore renewal property of a Poisson process gives $M_x - M_{x'} \sim \mathcal{P}(-\log(p_x/p_{x'}))$. Eventually one can conclude using the results of Proposition \ref{propo:stat proba estimator}.\qed
\end{proofprop}
\end{proposition}
We thus have defined an unbiased estimator for $m$.
\begin{remark} \label{rem:variance increase original ns}
As a matter of comparison, $\widetilde{m}$ can also be written $\widetilde{m} = \int_0^\infty \widetilde{p_x} \dx$. Then Remark \ref{rem:suboptimal p estimator} allows us to conclude that $\tilde{m}$ has a positive bias of order 1/N.
\end{remark}

\begin{proposition}[Finiteness of $\var{\m}$] \label{propo:finiteness var mk}
\[
\forall N \geq 2,\; \var{\m} \leq \dfrac{2}{1+1/N} E[X^{1+1/N}]^{2/(1+1/N)}.
\]
\begin{proofprop}
Starting from the expression of the variance found in Proposition \ref{propo_stat_estimateur}:
\[
\var{\m} = 2 \int_0^\infty  p_x \int_0^x p_{x'}^{1-1/N} \dx' \dx - \E{X}^2 ,
\]
we make use of Hölder's inequality:
\begin{multline*}
\int_0^x p_{x'}^{1-1/N} dx' \\
\leq \l \int_0^x dx'\r^{1/N} \l \int_0^x p_{x'}dx'\r^{1-1/N} \\
\leq x^{1/N} \left( \int_0^\infty p_{x'} \dx' \right)^{1-1/N}\\
\shoveright{\leq x^{1/N} \E{X}^{1-1/N} .}
\end{multline*}
And therefore:
\[
\var{\m} \leq \dfrac{2}{1+1/N} \E{X}^{1-1/N} \E{X^{1+1/N}} .
\]
Using Hölder's inequality again, one gets:
\[
\var{\m} \leq \dfrac{2}{1+1/N} \E{X^{1+1/N}}^{\frac{2}{1+1/N}} .
\] \qed
\end{proofprop}
\end{proposition}
\begin{corollary}[Value of $N$] \label{coro:finiteness var mk N min} Let $\epsilon > 0$, if $\E{X^{1+\epsilon}} < \infty$ then for any $N \geq 1/\epsilon $, $\m$ has a finite variance.
\end{corollary}
While the usual Monte Carlo estimator requires the finiteness of $\E{X^{2}}$ to have a finite variance, this estimator only requires the finiteness of a moment of order $1 + \varepsilon$. This is especially interesting when $X$ is heavy-tailed and this case is further investigated in Section \ref{s:Application to heavy-tailed random variables}.

\subsection{Comparison with classical Monte Carlo}

As the finiteness condition of the variance of $\m$ is much weaker than for a naive Monte Carlo estimator, one can expect a globally lower variance. This result is shown in Proposition \ref{propo_variance_comparison}. We first recall the crude Monte Carlo estimator:
\begin{equation} \label{eq:monte carlo definition}
\mmck \= \dfrac{1}{N}\sum \limits_{i = 1}^N X_i
\end{equation}
with $(X_i)_i$ $N$ \iid random variables with law $\mu^X$.
\begin{proposition}\label{propo_variance_comparison}
For any $N \geq 2$, $\var{\m} \leq \var{\mmck}$.
\begin{proofprop}
On the one hand one has:
\[
N \var{\mmck} +m^2 = 2 \int_0^\infty x p_x \dx ,
\]
and on the other hand one can write:
\[
N \var{\m} + m^2
=2 \int_0^\infty p_x \int_0^x p_{x'} \left[ N (p_{x'}^{-1/N}-1)+1 \right] dx' dx .
\]
Considering $f :p \mapsto p \left[ N (p^{-1/N}-1)+1 \right] $, we have $f(1)=1$ and:
\[
f'(p) = (N-1) (p^{-1/N} -1) \geq 0 \,, \forall p \in [0, 1].
\]
Thus: $\forall p \in [0,1], f(p) \leq 1$.
Therefore
\[
N \var{\m} + m^2 \leq 2 \I{0}{\infty} x p_x \dx
\]
which shows that $\var{\m} \leq \var{\mmck}$. \qed
\end{proofprop}
\end{proposition}
Thus the \emph{ideal} nested sampling estimator \eqref{eq:mk link nested} is always better than classical Monte Carlo in terms of variance and especially does not require the finiteness of the second-order moment of $X$ to have a finite variance.

\section{Randomised unbiased estimator} \label{section_practical_implementation}

The ideal estimator \eqref{eq:mk def} defined in Section \ref{section_moments_def} is not directly usable as it requires to simulate an infinite number of terms in sum \eqref{eq:mk def}. While the usual nested sampling implementations propose to stop the algorithm either after a given number of iterations, or according to some criterion estimated at each iteration, we propose a randomised unbiased estimator using recent results on paths simulation.

\subsection{Definition}

We are facing the issue of estimating $\E{\m}$ while it is not possible to generate such a $\m$ in a finite computer time. This problem is well identified in the field of Stochastic Differential Equations (SDE) where one often intends to compute the expectation of a path functional while only discrete-time approximations are available. Recently there have been two major breakthroughs that address this issue: first the Multilevel Monte Carlo (MLMC) method \citep{giles2008multilevel} has introduced the idea of combining \emph{intelligently} different biased estimators (levels of approximations) to speed up the convergence and reduce the bias; then \citet{mcleish2011ageneral} and \citet{rhee2013unbiased} have introduced a general approach to constructing unbiased estimator based on a family of biased ones. Basically in our context it randomises the number of simulated steps of the Markov chain, and slightly modifies the weights of the nested sampling to \emph{remove the bias} of the final estimator.

More precisely let us consider the truncated estimators $(\m_n)_{n\geq 1}$:
\begin{align*}
\m_n &= \I{0}{X_n} \left( 1 - \dfrac{1}{N} \right)^{M_x} \dx =\Sum[i = 0][n - 1] \left( X_{i+1} - X_i \right) \left( 1 - \dfrac{1}{N} \right)^i
\end{align*}
and $T$ a non-negative integer-valued random variable independent of $(X_n)_{n \in \N}$ such that $\forall i \in \N, \proba{T \geq i} \= \beta_i > 0$; one builds the following estimator (with $\m_0 = 0$):
\begin{align} \label{eq:definition zk}
\z &= \Sum[n = 0][\infty] \dfrac{\m_{n+1} - \m_{n}}{\proba{T \geq n}} \one_{T \geq n} = \Sum[n = 0][T] \dfrac{\m_{n+1} - \m_{n}}{\proba{T \geq n}} \notag \\
 &= \Sum[n=0][\infty] \left( X_{n+1} - X_n \right) \left( 1 - \dfrac{1}{N} \right)^n \dfrac{\one_{T \geq n}}{\proba{T \geq n}} .
\end{align}

\begin{remark}
The notation $\z$ might seem a bit confusing since $Z$ is used in the Introduction for the evidence as in \citep{skilling2006nested}. This is to keep consistency with \citet{rhee2013unbiased} notations where the randomising procedure comes from.
\end{remark}

\begin{proposition}[Statistical properties of $\z$] \label{propo:stat of zk}
\begin{align*}
\E{\z} &= m \\
\var{\z} &= \Sum[i=0][\infty] q_{i,N} \beta_{i}^{-1} - m^2
\end{align*}
with:
\begin{equation} \label{eq:definition qi}
q_{i,N} = 2 \l 1 - \dfrac{1}{N} \r^{2i} \I{0}{\infty} \I{x'}{\infty} p_x p_{x'}^{N-1} \dfrac{\lb -N \log p_{x'}\rb ^i}{i!} \dx \dx'  .
\end{equation}
\begin{proofprop}
Starting with the last formulation in \eqref{eq:definition zk} for $\zk$, one uses the fact that $T$ and $(X_i)_i$ are independent. Finally, \eqref{eq:mk def} and Proposition \ref{propo_stat_estimateur} let conclude: $\E{\z} = m$.

For the second-order moment, we use the fact that $\z$, like $\m$, can be written with an integral:
\[
\z = \I{0}{\infty} \l 1 - \dfrac{1}{N} \r ^{M_x} \dfrac{\one_{T \geq M_x}}{\proba{T \geq M_x}} \dx
\] and apply the same reasoning as for $\E{\m^2}$: given $x > x'$, the random variables $M_x - M_{x'}$, $M_{x'}$ and $T$ are independent, which brings:
\begin{multline*}
\E{\l 1 - \dfrac{1}{N} \r ^{M_x + M_{x'}} \dfrac{\one_{T \geq M_x}}{\proba{T \geq M_x}} \dfrac{\one_{T \geq M_{x'}}}{\proba{T \geq M_{x'}}}} \\
= \E{\l 1-\dfrac{1}{N}\r ^{M_x - M_{x'}} \l  1 - \dfrac{1}{N}\r ^{2 M_{x'}}  \beta_{M_{x'}}^{-1} \dfrac{ \one_{T \geq M_x}}{\proba{T \geq M_x}}} \\
= \E{\l 1-\dfrac{1}{N}\r ^{M_x - M_{x'}} \l  1 - \dfrac{1}{N}\r ^{2 M_{x'}}  \beta_{M_{x'}}^{-1}} \\
= \dfrac{p_x}{p_{x'}} \Sum[i=0][\infty] e^{N \log p_{x'}} \dfrac{\lb -N \log p_{x'} (1-1/N)^2 \rb ^i}{i!}\beta_{i}^{-1} \\
= \Sum[i=0][\infty] p_x p_{x'}^{N-1} \dfrac{\lb -N \log p_{x'} (1-1/N)^2 \rb ^i}{i!}\beta_{i}^{-1} .\\
\end{multline*}
Then using this equality in $\E{\z^2}$ gives the solution.\qed
\end{proofprop}
\end{proposition}
The asymptotic behaviour of the sequence $(q_{i,N})_i$ will drive the possible choices for the randomising distribution $(\beta_i)_i$: $\var{\Z}$ to remain finite implies that $q_{i, N} \beta_i^{-1} \to 0$ when $i \to \infty$.
\begin{lemma} \label{lemme:q decroit expo}
The sequence $(q_{i,N})_i$ goes to $0$ at least at exponential rate. Furthermore, if $X$ has density $f_X$ such that $\| f_X \|_\infty < \infty$, it is also bounded from below by an exponentially decreasing sequence.
\begin{prooflem}
Let $\varepsilon > 0$ be such that $\E{X^{1+\varepsilon}} < \infty$, $N \in \N \mid N > 1/\varepsilon$ and $i \geq 0$. We further extend the definition of $\var{\m}$ given in Proposition \ref{propo:stat proba estimator}, Eq. \eqref{eq:variance m} for any $N \in \R$. Proof of Proposition \ref{propo:finiteness var mk} is based on Hôlder's inequality and still holds in this case, and so for Corollary \ref{coro:finiteness var mk N min}. Hence, according to Corollary \ref{coro:finiteness var mk N min}: $ \exists N' \in \R$ such that $N' < N$ and $\var{\m}(N') < \infty$. Furthermore, given $x$ and $x'$ one can write:
\[p_x p_{x'}^{N-1}(-\log p_{x'})^i = p_x p_{x'}^{1-1/N'} p_{x'}^{N + 1/N' - 2}(-\log p_{x'})^i .
\]
Moreover the function $p : (0, 1) \mapsto p^{N + 1/N' - 2} (-\log p)^i$ is bounded above by $e^{-i} i^i (N + 1/N' - 2)^{-i}$. Using the Stirling lower bound $i \geq i^i e^{-i} \sqrt{2 \pi i}$ we can write:
\[p_x p_{x'}^{N-1}(-\log p_{x'})^i \leq p_x p_{x'}^{1-1/N'}\dfrac{i!}{\sqrt{2 \pi i} (N + 1/N' - 2)^i} .
\]
Finally, this inequality brings:
\[
q_{i,N} \leq \var{\m}(N') \l \dfrac{N(1-1/N)^2}{N + 1/N' -2} \r^i \dfrac{1}{\sqrt{2 \pi i}}
\]
and $(N + 1/N - 2)/(N + 1/N' -2)< 1$, which concludes the first part of the proof.

Let us now assume that $X$ has a density $f_X$. One has:
\begin{align*}
q_{i,N} &= 2 \I{0}{\infty} \I{0}{x} p_x p_{x'}^{N-1} \dfrac{\lb -N \log p_{x'} (1-1/N)^2 \rb ^i}{i!} \dx' \dx .
\end{align*}
Denote $x_L$ the left end point of $X$ (remember that $X$ is non-negative valued so $x_L \geq 0$). Then:
\begin{multline*}
q_{i,N} \geq 2 \I{x_L}{\infty} \I{x_L}{x} p_x p_{x'}^{N-1} \dfrac{\lb -N \log p_{x'} (1-1/N)^2 \rb ^i}{i!} \dx' \dx .
\end{multline*}

We then consider the change of variable $u = -\log p_x$ and $u' = -\log p_{x'}$; for all $i \geq 1$ one has:
\begin{align*}
q_{i,N} &\geq \dfrac{2}{\| f_X \|_\infty^2} \l 1 - \dfrac{1}{N} \r^{2 i} \I{0}{\infty} e^{-2 u} \I{0}{u} \dfrac{e^{-N u'} (N u')^i}{i!} \d u' \d u \\
&\geq \dfrac{2}{\| f_X \|_\infty^2} \l 1 - \dfrac{1}{N} \r^{2 i} \I{0}{\infty} e^{-2 u} \dfrac{1}{N} \Sum[k=i+1][\infty] \dfrac{e^{-N u} (N u)^k}{k!} \d u \\
&\geq \dfrac{2}{\| f_X \|_\infty^2} \dfrac{1}{N(N+2)} \l 1 - \dfrac{1}{N} \r^{2 i} \Sum[k=i+1][\infty] \l \dfrac{N}{N+2} \r^k \\
q_{i,N} &\geq \dfrac{1}{(N+2) \| f_X \|_\infty^2} \lb \dfrac{N}{N+2} \l 1 - \dfrac{1}{N} \r^2 \rb^i .
\end{align*}\qed
\end{prooflem}
\end{lemma}
Then it appears that the Geometric distribution plays a key role, as already stated by \citet{mcleish2011ageneral}. Hence we provide some theoretical results assuming that $T$ is a geometric random variable. 

\begin{proposition} \label{propo:var z_bar geom}
If $\proba{T \geq n} = e^{-\beta n},\; \beta > 0$, then:
\begin{equation} \label{eq:var z_bar geom}
\var{\z} = 2 \I{0}{\infty} \I{0}{x} p_x p_{x'}^{1-\frac{1}{\gamma(\beta,N)}}\dx' \dx - m^2
\end{equation}
with $\gamma(\beta,N) = N/(1 + (e^\beta-1)(N-1)^2)$.
\begin{proofprop}
Let $\alpha > 0$ be such that $(1-1/N) = e^{-\alpha}$. The argument is the same one as in Proposition \ref{propo:stat of zk}. One has:
\begin{multline*}
\E{\z^2} = 2 \I{0}{\infty} \I{0}{x} \E{ e^{-\alpha(M_x - M_{x'})} e^{(\beta - 2\alpha) M_x'}} \dx' \dx \\
= 2 \I{0}{\infty} \I{0}{x} p_x p_{x'}^{1 -\frac{1}{\gamma(\beta,N)}} \dx' \dx
\end{multline*}
with:
\begin{align*}
\dfrac{N}{\gamma(\beta, N)} &= 2N - N^2 + e^\beta (N - 1)^2 \\
&= 1 + (N-1)^2 (e^\beta - 1) .
\end{align*}\qed
\end{proofprop}
\end{proposition}
This expression is indeed the same as the one of Proposition \ref{propo_stat_estimateur} with the function $\gamma(\beta,N)$ instead of $N$. Hence the greater $\gamma$ the smaller $\var{\Z}$. Furthermore one has directly all the results from Section \ref{ss:Definition of the moment estimator}, especially the finiteness conditions for the variance given in Proposition \ref{propo:finiteness var mk} and Corollary \ref{coro:finiteness var mk N min}, replacing $N$ by $\gamma(\beta,N)$.

While there is no value of $\beta$ minimising $\var{\z}$ at a given $N$ (the smaller $\beta$ the smaller the variance of the randomised estimator $\z$), there is an optimal value for $N$ at a given $\beta$, $\ie$ for a given finite computational budget: $N = \sqrt{1 + \E{T}}$. One can reverse this relation, which gives:
\begin{equation} \label{eq:beta approximation}
\beta_\text{app} \= \log \l 1 + 1/(N^2-1) \r .
\end{equation}
\begin{corollary} \label{coro:double variance Z bar}
Let $N \geq 2$ and $\proba{T \geq n} = e^{-n \, \beta_\text{app}(N)}$, then:
\begin{equation}
\var{\Z}(N) = \var{\m} (\tfrac{N+1}{2}) \approx 2 \var{\m}(N).
\end{equation}
\begin{proofcoro}
Noticing that for any $N \geq 2$, one has $\gamma(\beta_\text{app}(N), N) = (N+1)/2$ gives the first equality. Then, since $\var{\m}$ typically scales with $1/N$ (usual results on nested sampling) gives the approximation.
\end{proofcoro}
\end{corollary}
This means that instead of choosing an arbitrary stopping criterion for nested sampling, randomising the number of iterations and computing $\Z$ allows for keeping an unbiased estimator without increasing drastically the variance (factor up to 2, reached with suboptimal implementation of Corollary \ref{coro:double variance Z bar}). This result will be illustrated in the examples of Section \ref{s:Example}.

\subsection{Convergence rate}
Throughout the paper we consider that the computational cost for generating a realisation of $\z$ is the number of simulated samples. Accordingly, in this section it is the number of calls to a simulator of a conditional law.

\begin{proposition} \label{propo_tau}
Let $\tau$ be the random variable of the number of samples required to generate $\z$. One has $\tau = N + T$.
\begin{proofprop}
If $T = 0$ then no other simulation is done other than the first element of each Markov chain, $\ie$ $N$ simulations are done. Then each step requires the simulation of the next stopping time, $\ie$ one simulation. Finally, this brings $\tau = N + T$.
\end{proofprop}
\end{proposition}

\begin{corollary}[Convergence rate of $\z$] \label{coro:convergence rate zk}
For any non-negative integer-valued randomising variable $T$ such that $\E{T} < \infty$ and $\forall i \in \N,\; \proba{T \geq i} > 0$, one has:
\begin{equation} \label{eq:order of magnitude convergence rate Z_bar}
\E{\tau} \cdot \var{\z} \geq 2 q_{1,2} + O \l \dfrac{1}{N} \r ,\; N \to \infty .
\end{equation}
\begin{proofcoro}
Note that $\var{\m} = \Sum[i=0][\infty] q_{i,N} - m^2$. Hence, one has $\var{\z} > \var{\m}$ because $\var{\z} = \var{\m} \Leftrightarrow \forall i \in \N,\; \beta_i = \proba{T \geq i} = 1$ and $\E{\tau} > N$ because $\E{\tau} = N \Leftrightarrow \E{T} = 0$ while $\forall i \in \N,\; \proba{T \geq i} > 0$. Furthermore, the power series expansion of the exponential function and the dominated convergence theorem let us rewrite $\var{\m}$:
\begin{align*}
\var{\m} &= \Sum[i=1][\infty] 2 \I{0}{\infty} \I{x'}{\infty} p_x p_{x'} \dfrac{(-\log p_{x'})^i}{N^i i!} \dx \dx' \\
\var{\m} &= \Sum[i=1][\infty] q_{i,2} \l \dfrac{2}{N} \r^i
\end{align*}
which brings: $\var{\m} = q_{1,2} \cdot 2/N + O \l 1/N^2 \r$. All together, these inequalities complete the proof.\qed
\end{proofcoro}
\end{corollary}

If the inequality \eqref{eq:order of magnitude convergence rate Z_bar} is close to an equality then $\z$ has a canonical square-root convergence rate (as a function of the computational cost). However there is no guarantee on this rate of convergence. Especially Corollary \ref{coro:order of magnitude of var alpha geom} below shows that it is not the case when $T$ has a geometric distribution.

\begin{corollary} \label{coro:order of magnitude of var alpha geom}
If $T$ is a Geometric random variable such that $\forall n \in \N, \; \proba{T \geq n} = e^{-\beta n}$ with $\beta = \Theta(1/N^{1+\varepsilon})$, $\varepsilon \geq 0$, then:
\[
\begin{cases}
\E{\tau} \cdot \var{\z} = \Theta \l N \r & \varepsilon \in [0, 1] \\
\E{\tau} \cdot \var{\z} = \Theta \l N^{\varepsilon} \r & \varepsilon > 1 .
\end{cases}
\]
\begin{proofcoro}
Denote $B = 1/(e^\beta - 1)$; one has:
\[
\dfrac{N+B}{\gamma(B,N)} = N + \dfrac{B}{N} + \dfrac{N^2}{B} - 1 - \dfrac{2N}{B} + \dfrac{1}{B} + \dfrac{1}{N} .
\]
With $\beta = \Theta(1/N^{1+\varepsilon})$, $\varepsilon \geq 0$, one has $B \sim 1/\beta \sim N^{1+\varepsilon}$. Finally, this gives:
\[
\dfrac{N+B}{\gamma(B,N)} \sim N + N^\varepsilon + N^{1-\varepsilon} + O(1),
\]
which concludes the proof.\qed
\end{proofcoro}
\end{corollary}

Hence the unbiased randomised estimator of Corollary \ref{coro:double variance Z bar} with $\beta = \beta_\text{app} = \Theta(1/N^2)$ does not have a canonical square-root convergence rate. Furthermore, even though the realisation of the geometric random variable gives a small number of iterations, one may want to run the algorithm longer to probe the tail of the likelihood function to make sure that no important part is missing \citep{skilling2006nested}. That is why the idea behind randomised estimators is to average several replicas of $\z$ because it will somehow average the quantities $\one_{T \geq n} / \proba{T \geq n}$ in \eqref{eq:definition zk}. More precisely, let $G(c)$ be the random variable of the number of simulations of $\z$ one can afford with a computational budget $c$:
\[
G(c) = \max \{ n \geq 0 \mid \sum_{i=1}^n \tau_i \leq c \}
\]
where $\tau_i$ is the computational effort required to generate the $i^{th}$-sample $\z_i$, one considers the following estimator:
\begin{equation} \label{eq:definition alpha}
\alp(c) = \dfrac{1}{G(c)} \sum \limits_{i=1}^{G(c)} \z_i .
\end{equation}
In this setting \citet{glynn1992asymptotic} showed a CLT-like result:
\begin{equation} \label{eq:CLT alpha}
c^{1/2}(\alp(c) - \E{\z}) \xrightarrow[c \rightarrow \infty]{\mathcal{L}} (\E{\tau} \cdot \var{\z})^{1/2} \mathcal{N}(0,1) .
\end{equation}
Hence in our context one has to tune $(\beta_i)_i$ and $N$ to minimise the product $\E{\tau} \cdot \var{\z}$.

\subsection{Optimal randomisation}
Since $T$ is a non-negative random variable one has $\proba{T \geq 0} = \beta_0 = 1$. Let $\mathcal{C} = \{ (\beta_i)_{i} \in (0,1]^\N \mid \beta_0 = 1 \text{ and } \forall i \in \N \,, \beta_{i+1} \leq \beta_{i} \}$; we intend to solve the optimisation problem:
\begin{multline} \label{eq:optim problem}
\underset{\substack{ (\beta_i)_i \in \mathcal{C} \\ N \in \llbracket 2, \infty ) }}{\argmin} \E{\tau} \cdot \var{\z} \\
= \underset{\substack{ (\beta_i)_i \in \mathcal{C} \\ N \in \llbracket 2, \infty ) }}{\argmin} \l N-1 + \Sum[i=0][\infty] \beta_i \r \l  \Sum[i=0][\infty] q_{i,N} \beta_i^{-1} - m^2\r 
\end{multline}
where the $(q_{i,N})_i$ are given by \eqref{eq:definition qi}. Furthermore, one can rewrite the $(q_{i,N})_i$ assuming that $X$ has a density $f_X > 0$. Indeed in this context $X_n$ has a density $f_n$ such that:
\begin{equation*}
\forall n \geq 1 ,\; f_n(x) = N \dfrac{p_x^{N-1} (-N \log p_x)^{n-1}}{(n-1)!} f_X(x) .
\end{equation*}
This gives:
\[
\forall i \in \N ,\; q_{i,N} = \dfrac{2}{N} \l 1 - \dfrac{1}{N} \r^{2i} \E{\mathcal{R}(X_{i+1})}
\]
with $\mathcal{R}(x) = \int_x^\infty p_u \d u/f_X(x)$. Hence we further assume that $(q_{i,N})_i$ is decreasing, which is the case for a Pareto random variable (see Section \ref{ss:Exact resolution for a Pareto distribution}) and at least for any distribution for which $\mathcal{R}$ is non-increasing like exponential and uniform distributions. In this context Proposition \ref{propo:optimal distribution for T} gives the optimal distribution for $T$ for a given $N$.

\begin{proposition}[Optimal distribution for $T$] \label{propo:optimal distribution for T}
If $(q_{i,N})_{i \geq 1}$ is decreasing then the optimal distribution $(\beta_i^*)_i$ for $T$ is given by:
\begin{align*}
\forall i \in \llbracket 0, i_0 \rrbracket &\,, \beta_i^* = 1 \\
\forall i > i_0 &\,, \beta_i^* = \sqrt{\dfrac{N + i_0}{S_0}}\sqrt{q_{i,N}}
\end{align*}
with $i_0 = \min \{ i \in \N \mid \sum_{j=0}^{i} q_{j,N} - m^2 > (N + i)q_{(i + 1),N} \}$ and $S_0 = \sum_{j = 0}^{i_0} q_{j,N} - m^2$.
\begin{proofprop}

First one shows that $i_0$ is well determined. The sequence $(\Delta_i)_i$ defined by:
\[\forall i \in \N \,, \Delta_i = \sum_{j=0}^i q_{j,N} - m^2 - (N+i)q_{(i+1),N}\]
is increasing:
\begin{align*}
\Delta_{i+1} - \Delta_i &= q_{(i+1),N} - (N + i + 1)q_{(i+2),N} + (N+i) q_{(i+1),N} \\
&= (N + i + 1) (q_{(i+1),N} - q_{(i+2),N}) > 0 .
\end{align*}
Furthermore $q_0 - m^2 = 2 \I{0}{\infty} \I{x'}{\infty} p_x p_{x'} \l p_{x'}^{N-2} - 1 \r \dx \dx' \leq 0 < N q_{1,N}$, so $\Delta_0 < 0$, and $\Delta_i \to \var{\m}$ when $i \to \infty$ because $(q_{i,N})_i$ decreases at exponential rate. So there exists $i_0 \in \N \mid \Delta_{i_0 - 1} \leq 0 \text{ and } \Delta_{i_0} > 0$.

Let us now consider the auxiliary problem:
\[
\underset{\substack{(\beta_i)_{i \geq 1} \\ \beta_i > 0}}{\argmin} \l \beta + \Sum[i=1][\infty] \beta_i \r \l q + \Sum[i=1][\infty] q_{i,N} \beta_i^{-1} \r
\]
with $\beta > 0$ and $q \in \R$. We show that it has a solution if and only if $q > 0$. Let $i \geq 1$, cancelling the partial derivatives brings:
\[
\forall i \geq 1 ,\; 0 = \l q + \Sum[j=1][\infty] q_j \beta_j^{-1} \r + \l \beta + \Sum[j=1][\infty] \beta_j \r \dfrac{-q_{i,N}}{\beta_i^2} .
\]
Then the solution should be of the form: $\forall i \in \llbracket 1, \infty) \,, \beta_i = c_0 \sqrt{q_i}$ for some $c_0 > 0$. Solving now the problem with $c_0$, the derivative writes $q - \beta/c_0^2$. If $q \leq 0$ then it is strictly decreasing and there is no global minimiser. On the contrary, $q > 0$ brings $c_0 = \sqrt{\beta/q}$ and $\forall i \geq 1 \,, \beta_i = c_0 \sqrt{q_i}$.

Thus, in our context with the constraint $\forall i \in \N \,, \beta_i \leq 1$, this means that solving the optimisation problem will set iteratively $\beta_i = 1$ until the minimiser is feasible, $\ie$ until $i_0 \= \min \{i \in \N \mid \Sum[j=0][i] q_{j,N} - m^2 > (N+i) q_{(i+1),N} \}$. Then the solution will be given by:
\begin{align*}
\forall i \in \llbracket 1, i_0 \rrbracket &\,, \beta_i = 1 \\
\forall i > i_0 &\,, \beta_i = \dfrac{\sqrt{q_{i,N}}}{\sqrt{\dfrac{1}{N + i_0} \Sum[j=0][i_0] (q_{j,N} - m^2)}} .
\end{align*}\qed
\end{proofprop}
\end{proposition}
It is part of the proof in the appendix that $i_0$ is well defined and so it appears that the optimal distribution enforces the estimator to go at least until the $i_0^{th}$ event. Recalling $(X_n)_n$ is the cumulated Markov Chain (associated with the marked Poisson Process with parameter $N$), this can be understood in the sense that on average, at least $N$ events are necessary to \emph{use} at least one time each process. Even if the link between $i_0$ and $N$ is not that straightforward, one can then conjecture that $\lim \limits_{N \to \infty} i_0 = \infty$.

\begin{corollary}[Bounds on $\beta_i^*$]
For all $i > i_0$, one has:
\begin{equation}
\sqrt{\dfrac{q_{i,N}}{q_{i_0 + 1,N}}} > \beta_i^* \geq \sqrt{\dfrac{q_{i,N}}{q_{i_0,N}}} .
\end{equation}
\begin{proofcoro}
By definition of $i_0$, one has:
\[
(N + i_0) q_{i_0 + 1} < \Sum[j=0][i_0] q_j - m^2 \leq (N+i_0 - 1) q_{i_0} + q_{i_0}
\]
which concludes the proof.\qed
\end{proofcoro}
\end{corollary}
Thus the tail of the optimal distribution $(\beta_i^*)_i$ is exponentially decreasing by Lemma \ref{lemme:q decroit expo}. From these bounds on the $(\beta_i)_i$ one can also derive bounds on the variance:
\[
q_{i_0 + 1,N} \E{\tau}^2 < \E{\tau} \cdot \var{\z} \leq q_{i_0,N} \E{\tau}^2 .
\]
Assuming $\lim_{N \to \infty} i_0 = \infty$ and using the lower bound on $q_{i,N}$ from Lemma \ref{lemme:q decroit expo}, one can show that $\lim_{N \to \infty} \E{\tau} \cdot \var{\z} = \infty$, which implies the existence of an optimal $N$. Section \ref{ss:Exact resolution for a Pareto distribution} presents an exact resolution of this optimisation problem for a Pareto random variable.

Finally, we have presented in this section the framework for an optimal resolution of Problem \eqref{eq:optim problem} and proven existence of a solution under reasonable assumptions ($(q_{i,N})_i$ is decreasing and $\lim_{N \to \infty} i_0 = \infty$). Furthermore the comprehensive resolution in the case of a Pareto distribution in Section \ref{ss:Exact resolution for a Pareto distribution} legitimises these assumptions. Generally speaking, if $(q_{i,N})_{i\geq 1}$ is not decreasing the optimisation has to be performed over all the decreasing sub-sequences of $(q_{i,N})_i$, which turns it into a combinatorial problem \citep[see][Theorem 3]{rhee2013unbiased}.

\subsection{Geometric randomisation}
On the one hand the computation of the optimal distribution for $T$ can be quite demanding in computer time; and on the other hand the geometric law plays a key role as for any distribution $p_x$, the sequence $(q_{i,N})_i$ decreases at exponential rate and the optimal randomising distribution (when $(q_{i,N})_i$ is decreasing) is somehow a shifted geometric law. Therefore we study the parametric case where $\proba{T \geq n} = e^{-\beta n}$, $\beta > 0$ and tune $\beta$ and $N$ to minimise $\E{\tau} \cdot \var{\z}$.

Using the exponential power series in $\var{\z}$ (cf. Eq. \eqref{eq:var z_bar geom}), the optimisation problem \eqref{eq:optim problem} becomes:
\begin{equation} \label{eq:problem optim geom}
\underset{\substack{ \beta > 0 \\ N \in \llbracket 2, \infty ) }}{\min} \l N + \dfrac{1}{e^{\beta} - 1} \r \l \Sum[i=0][\infty] q_{i,2} \l \dfrac{2}{\gamma(\beta,N)} \r^i - m^2 \r .
\end{equation}

\begin{proposition} \label{propo:existence solution cas parametrique}
There exists a global minimiser $(\beta_\text{opt}, N_\text{opt})$ to Problem \eqref{eq:problem optim geom}. Furthermore, $(\beta_\text{opt}, N_\text{opt})$ satisfies the relationship:
\begin{equation} \label{eq:relation beta_opt N_opt}
\beta_\text{opt} = \log \l 1 + \dfrac{2}{N_\text{opt}^2 - 1 + (N_\text{opt}-1) \sqrt{N_\text{opt}^2 + 6 N_\text{opt} + 1}} \r .
\end{equation}
\begin{proofprop}
Denote: \[Q_N(\beta) = \l N + \dfrac{1}{e^\beta - 1} \r \l \Sum[i=0][\infty] q_{i,2} (2/\gamma)^i - m^2 \r \] the quantity one seeks to minimise.

First, we show that for any fixed $N$, there exists a global minimiser of $Q_N(\beta)$. One has $Q_N(\beta) \to \infty$ when $\beta \to 0$ and $\gamma(\beta,N) \to 0$ when $\beta \to \infty$. Hence, either $\exists \beta_\infty \in (0, \infty]$ such that:
\[
\begin{cases}
Q_N(\beta) \xrightarrow[\beta \nearrow \beta_\infty]{} \infty \\
Q_N(\beta) < \infty & \forall \beta < \beta_\infty .
\end{cases}
\]
Then $Q_N$ is continuous on $(0, \beta_\infty)$ with infinite limits on $0$ and $\beta_\infty$, so it reaches its minimum on $(0, \beta_\infty)$; or $\exists \beta_\infty \in (0, \infty)$ such that:
\[
\begin{cases}
Q_N(\beta) < \infty & \forall \beta \in (0, \beta_\infty] \\
Q_N(\beta) = \infty & \forall \beta > \beta_\infty .
\end{cases}
\]
Since $Q_N$ is continuous on $\beta_\infty^-$ by Monotone Convergence Theorem, $Q_N$ reaches its minimum on $(0, \beta_\infty]$.

Let $\beta_\text{opt}(N) > 0$ be such that $\inf_\beta Q_N(\beta) = Q_N(\beta_\text{opt})$. We now show that there exists an optimal $N$. It is sufficient to show $Q_N(\beta_\text{opt}) \to \infty$ when $N \to \infty$. Denote $B = 1/(e^\beta - 1)$; one has:
\[
\dfrac{1}{\gamma(B,N)} = \dfrac{1}{N} + \dfrac{N}{B} - \dfrac{2}{B} + \dfrac{1}{NB} .
\]
Hence, depending on the growth rate of $B$ when $N \to \infty$, one would have:
\begin{align*}
B = O \l N \r, & \dfrac{1}{\gamma} \sim  \dfrac{N}{B} \Rightarrow \inf_\beta Q_N(\beta) \xrightarrow[N \to \infty]{} \infty \\
N = o \l B \r, & \dfrac{1}{\gamma} \sim \dfrac{1}{N} \text{ or } \dfrac{N}{B} \Rightarrow \inf_\beta Q_N(\beta) \sim \dfrac{B}{N} \text{ or } N \\
& \Rightarrow \inf_\beta Q_N(\beta) \xrightarrow[N \to \infty]{} \infty .
\end{align*}
Then in any cases $Q_N(\beta_\text{opt}) \to \infty$ when $N \to \infty$, which means that there exists $N_\text{opt} \in \N \mid Q_{N_\text{opt}}(\beta_\text{opt}) = \inf_N Q_N(\beta_\text{opt})$.

We now show the relationship between $\beta_\text{opt}$ and $N_\text{opt}$: the partial derivatives of $\E{\tau} \cdot \var{\zk}$ against $B$ and $N$ write:
\[
\begin{cases}
\dfrac{\partial \l \E{\tau} \cdot \var{\zk} \r}{\partial B} = \var{\zk} + \E{\tau} \dfrac{\partial \var{\zk}}{\partial \gamma} \dfrac{\partial \gamma}{\partial B} \\
\dfrac{\partial \l \E{\tau} \cdot \var{\zk} \r}{\partial N} = \var{\zk} + \E{\tau} \dfrac{\partial \var{\zk}}{\partial \gamma} \dfrac{\partial \gamma}{\partial N} .\\
\end{cases}
\]
At point $(\beta_\text{opt}, N_\text{opt})$, both equations are cancelled, which gives:
\[
\dfrac{\partial \gamma}{\partial N}(B_\text{opt}, N_\text{opt}) = \dfrac{\partial \gamma}{\partial B}(B_\text{opt}, N_\text{opt}) .
\]
Recalling $\gamma(B,N) = N B /(B + (N-1)^2)$, this gives the equation: $B_\text{opt}^2 - (N_\text{opt}^2-1)B_\text{opt} - N_\text{opt}(N_\text{opt}-1)^2 = 0$. One can solve it in $B_\text{opt}$ and keep the positive root, which gives the solution.
\end{proofprop}
\end{proposition}

Hence there is always an optimal solution to Problem \eqref{eq:problem optim geom}, meaning this parametrisation is meaningful.

To summarise we have shown that by randomising the finite number of iterations and slightly modifying the weights of the original nested sampling, it is possible to define an unbiased estimator for the mean of any real-valued random variable with continuous \cdf, resolving the issue of choosing an \emph{appropriate} stopping criterion. With a suboptimal geometric randomisation as in Corollary \ref{coro:double variance Z bar}, the variance is at most twice the one of the ideal case (estimator \eqref{eq:mk def}). However it is not usable with a fixed predetermined computational budget and its convergence rate is slower than the canonical square-root one. To circumvent this limitation, the idea is to average several replicas of the randomised unbiased estimator (see Eq. \eqref{eq:definition alpha}). This new estimator remains unbiased and also supports a Central Limit Theorem.

All these theoretical results assume that it is possible to generate conditional random variables when required, as for the original nested sampling algorithm \citep[see][Section 9]{skilling2006nested}. Efficient conditional simulation can be carried out in different ways, from perfect simulation \citep[see for example][]{propp1996exact} to approximation using random walk Metropolis-Hastings. The aim of this paper is not to challenge this hypothesis in a general manner but only to provide a new insight on the risk of choosing a \emph{bad} stopping criterion in nested sampling, and to propose an other tool to deal with this issue. Since nested sampling has been applied successfully to a great number of problems so far, these results are expected to hold in these situations. Also the examples of Section \ref{s:Example} are in good agreement with these theoretical results.

In the next section, we discuss the different stopping criteria usually recommended for nested sampling and parallel implementation of the estimators.

\subsection{Parallel implementation}

\citet[][Section 7]{skilling2006nested} presents two possible termination rules based on criteria evaluated on-the-fly:
\begin{itemize}
\item stop when the greatest expected increment (current weight and biggest found likelihood value) is smaller than a given fraction of the current estimate;
\item stop when the number of iterations \emph{significantly} exceeds $N H$ with $H$ the information, estimated on-the-fly.
\end{itemize}
\citet{chopin2010properties} use an other stopping criterion, close to the first one above, it is: ``stop when the new increment is smaller than a given fraction of the current estimate". An other option is to do a predetermined number of iterations \citep{brewer2011diffusive}. Unfortunately these criteria give no guarantee on the convergence of the estimator to the sought value and may lead to biased estimation.

A first difference between the three first criteria and the last one stands in the fact that this latter uses a known computational budget while the others ones will run until the criterion is satisfied; hence there is no way to estimate the (random) final number of iteration in advance. This difference is also to be found between $\Z$ and $\alp$: the first one will use a random number of simulated samples (the draw of the randomising variable) while the second one is defined with a fixed computational budget. Hence these two categories of estimators cannot be compared because the setting is not the same.

An other main difference between these estimators is whether they enable parallel computation or not. The three first stopping criteria need to be evaluated at each iteration and are based on quantities estimated with the full process with parameter $N$. Hence they do not allow for parallel computation. On the other hand, with a predetermined total number of iterations, parallel computation on the model of \citep[][Section 4.2]{walter2015moving} can be carried out. The randomised estimator $\Z$ also enables this feature as the random number of iterations is drawn before the algorithm starts. Considering $\alp$, each replica can be computed in parallel, and further the computation of each replica also allows for parallel implementation. Hence $\alp$ allows for a \emph{double} parallelisation, which is worth noticing as it may require a substantial computational budget to become effectively Gaussian.

To conclude, one stresses out the fact that among estimators with random computational budget, $\Z$ is the only one allowing for parallel computation; furthermore it is also the only one unbiased and its variance is at worst twice the one of the ideal estimator (upper bound reached with suboptimal implementation of $\Z$ as in Corollary \ref{coro:double variance Z bar}). Both fixed-budget estimators enable parallel implementation; however nested sampling with a predetermined number of iterations has no reason to be close to the sought value. On the other hand, $\alp$ is unbiased and supports a CLT. All these considerations are illustrated in Section \ref{s:Example}.

\section{Application to heavy-tailed random variables}
In this section we give insights on the properties of the new estimator when $X = g(\U)$ is heavy-tailed. Mean estimation for heavy-tailed random variables is a well identified problem often addressed by some parametric assumptions on the \cdf{} of $X$; see \citet{beirlant2012overview} for a comprehensive overview of tail index estimation, and \citet{peng2001estimating,johansson2003estimating,necir2010estimating,hill2013robust} for references on mean estimation for heavy-tailed random variables.

In the sequel we then give explicit results for the Pareto distribution $p_x = \proba{X > x} = 1 \wedge x^{-a} ,\; a > 1$.

\subsection{Exact resolution for a Pareto distribution}
With an analytic form for the \cdf{} of $X$, we can derive explicit formulae for the variance (Eq. \eqref{eq:variance m}) and the optimisation problem \eqref{eq:optim problem}.

First we compare the variance of the ideal estimator $\m$ against usual Monte Carlo and Importance Sampling estimators. In this latter case the importance density is chosen to be a Pareto distribution with parameter $b > 0$.
\begin{proposition}[Variance comparison] \label{propo:var_pareto}
For a Pareto distribution, one has $m = a/(a-1)$ and the variances write:
\begin{align*}
a > 2 ,\; & \var{\mmck} = \dfrac{m (m-1)^2}{2N - m N}\\
a > \dfrac{2N}{2N-1} ,\; &\var{\m} = \dfrac{m (m-1)^2}{2 N - m} \\
a > 1 + \dfrac{b}{2} ,\; &\var{\mis} = \dfrac{m^2(B-1)^2}{N (2B - 1)}
\end{align*}
with $B = (a-1)/b \in (1/2, \infty)$.
\begin{proofprop}
For the first equality:
\begin{align*}
\E{X} &= \int_0^\infty p_x \dx = \dfrac{a}{a-1} \\
\var{\mmck} &= \dfrac{1}{N} \left( \E{X^{2}} - \E{X}^2 \right) = \dfrac{a}{N(a-2)(a-1)^2} \\
&= \dfrac{m (m - 1)^2}{(2-m)N} ;
\end{align*}
for the second one:
\begin{align*}
\E{\m^2} &= 2 \int_0^\infty \int_0^x p_x p_{x'}^{1-1/N} \dx' \dx \\
&= 2 \int_0^1 \int_0^x \cdots + 2 \int_1^\infty \int_0^1 \cdots + 2 \int_1^\infty \int_1^x \cdots \\
&= 1 +\dfrac{2}{a-1} + \dfrac{2}{(a-1)(2(a-1) - a/N)}\\
\var{\m} &= \dfrac{a}{N(a-1)^2 (2(a-1) - a/N)} ;
\end{align*}
and for the third one:
\begin{align*}
\var{\mis} &= \dfrac{1}{N} \lb  \int_1^\infty x^{2} \dfrac{a^2}{b} x^{-2a + b -1} \dx - \dfrac{a^2}{(a-1)^2} \rb \\
\var{\mis} &= \dfrac{a^2}{N(a-1)^2} \l \dfrac{1}{B(2-B)} - 1 \r 
\end{align*}
with $B = b/(a-1)$.\qed
\end{proofprop}
\end{proposition}

It is clearly visible that the classical Monte Carlo estimator needs a second-order moment while $\m$ only requires $a > 2N/(2N-1) \approx 1 + 1/2N$ and $\mis$ requires $a > 1 + b/2$; it also illustrates the result of Proposition \ref{propo_variance_comparison}: $\var{\m} < \var{\mmck}$. The optimal value $b = a-1$ cancels out $\var{\mis}$. It is well known that there is an optimal density $q$ for IS that cancels out the variance of the IS estimator but it is case-specific: here a Pareto density with parameter $a-1$.

\begin{remark}[Limit distribution of classical Monte Carlo estimator] \label{rem:limit distribution MC}
In the case of Pareto distribution, when $a > 2$ the Central Limit Theorem gives the limit law of the estimator while for $1 < a < 2$ the Generalised Central Limit Theorem \citep[see for example][]{embrechts1997modelling} states that $\sum_i X_i$ is in the domain of attraction of a stable law with parameter $a$:
\[
N^{1-1/a} \l \dfrac{1}{N} \Sum[i=1][N] X_i - m \r  \dfrac{1}{C_a} \xrightarrow[N \to \infty]{\mathcal{L}} X_a
\]
with the characteristic function of $X_a$, $\phi_{X_a}$, writing $\phi_{X_a}(t) = \exp \lb  -\vert t \vert^a \l 1-i \l \tan \l\pi a /2 \r  \r  sgn(t) \r  \rb $ and $C_a$ the normalising constant \(
C_a = \pi^{1/a} \l 2 \Gamma(a) \sin \pi a / 2 \r ^{-1/a}
\).
\end{remark}

We now detail the resolution of optimisation problems \eqref{eq:optim problem} and \eqref{eq:problem optim geom}. Especially we first explicit the form of the sequence $(q_{i,N})_i$ defined in Eq. \eqref{eq:definition qi}.

\begin{proposition} \label{prop:beta opt pareto}
If $X$ is a Pareto random variable with parameter $a > 1$, then:
\begin{equation*}
\forall i \in \N \,, q_{i,N} = \dfrac{2}{(a-1)(aN - 2)} \lb \dfrac{a(N-1)^2}{N(aN-2)} \rb ^i + \dfrac{\one_{i=0}(a+1)}{2(a-1)}.
\end{equation*}
\begin{proofprop}
Let $i \geq 0$, one has:
\begin{multline*}
\I{1}{\infty} \I{x'}{\infty} p_x p_{x'}^{N-1} \dfrac{\lb  -N \log p_{x'} (1-1/N)^2 \rb ^i}{i!} \dx \dx' =\\
\dfrac{\lb aN(1-1/N)^2 \rb ^i}{i!} \I{1}{\infty} \I{x'}{\infty} x^{-a} x'^{-a(N-1)} (\log x')^i \dx \dx'\\
= \dfrac{\lb aN(1-1/N)^2 \rb ^i}{(a-1) i!} \I{1}{\infty} x'^{1-aN} (\log x')^i \dx' \\
= \dfrac{\lb aN(1-1/N)^2 \rb ^i}{(a-1) i!} \dfrac{\Gamma(i+1)}{(aN - 2)^{i+1}} \\
\shoveright{= \dfrac{1}{(a-1)(aN-2)} \lb  \dfrac{aN}{aN-2} \l 1-\dfrac{1}{N} \r ^2 \rb ^i}
\end{multline*}
with $\Gamma$ standing here for the Gamma function. Furthermore:
\begin{multline*}
\I{0}{1} \I{x'}{\infty} p_x p_{x'}^{N-1} \dfrac{\lb  -N \log p_{x'} (1-1/N)^2 \rb ^i}{i!} \dx \dx' = \dfrac{\one_{i=0}(a+1)}{2(a-1)} .\\
\end{multline*}
$(q_{i,N})_i$ is decreasing iff:
\[
\dfrac{aN}{aN - 2} \l 1-\dfrac{1}{N} \r^2 < 1 \Leftrightarrow 1 < a \l 1 - \dfrac{1}{2N} \r
\]
which is indeed the condition for the finiteness of $\var{\m}$ already stated in Proposition \ref{propo:var_pareto}.\qed
\end{proofprop}
\end{proposition}
Hence for a Pareto distribution $(q_{i,N})_i$ is decreasing. One can then look for $i_0$, the solution of the problem $i_0 = \min \{ i \in \N \mid \sum_{j=0}^{i} q_{j,N} - m^2 > (N + i)q_{(i + 1),N} \}$.
Whilst an exact solution can be expressed using the lower branch of the Lambert W function \citep[see for example][]{corless1996lambertw}, the following proposition gives an asymptotic approximation when $N \to \infty$ to precise the growth rate of $i_0$.
\begin{proposition}
If $X$ is a Pareto random variable, then:
\begin{equation*}
i_0 = \dfrac{N m}{2} \l \log N + \log \log N - \log( \dfrac{m}{2} ) \r + o(N),\; N \to \infty.
\end{equation*}
\begin{proofprop}
The problem can be rewritten:
\[
\min \left\lbrace i \geq 1 \mid \dfrac{1}{1-\beta} - \dfrac{aN - 2}{2(a-1)} > \beta^{i+1} \l N + i + \dfrac{1}{1 - \beta} \r \right\rbrace .
\]
Furthermore one has:
\[
\dfrac{1}{1-\beta} = \dfrac{N m}{2} + \dfrac{(a-2)^2}{4(a-1)^2} + o(1)
\]
which brings that the left hand term is equal to $(m / 2)^2 + o(1)$. Writing $i = N(k_0 + k_1 \log N + k_2 \log \log N)$ brings:
\[
\beta^{i+1} = e^{-\frac{2 k_0 }{m}} N^{-\frac{2 k_1}{m}} \l \log N \r^{-\frac{2 k_2}{m}} \l 1 + o(1) \r .
\]
Hence one has to choose $k_0$, $k_1$ and $k_2$ such that the right hand term also equals $(m/2)^2 + o(1)$, which gives the solution.\qed
\end{proofprop}
\end{proposition}

\begin{corollary}[Order of magnitude of $\E{\tau} \cdot \var{\zk}$] \label{coro:order of magnitude var alpha}
\begin{equation*}
\E{\tau} \cdot \var{\z} \underset{N \to \infty}{\sim} \l \dfrac{m(m - 1)}{2} \r^2 \log N .
\end{equation*}
\begin{proofcoro}
Using the asymptotic expansion of $i_0$ one finds $q_{i_0} \sim (N^2 \log N )^{-1} (m - 1)^2$. Furthermore, one has $\E{\tau} \sim i_0$. Finally, the use of $\E{\tau} \cdot \var{\z} \sim q_{i_0} \E{\tau}^2$ gives the result.\qed
\end{proofcoro}
\end{corollary}
Corollary \ref{coro:order of magnitude var alpha} shows that $\E{\tau} \cdot \var{\z} \to \infty$ when $N \to \infty$ so there is an optimal value for $N$ that minimises $\E{\tau} \cdot \var{\Z}$; a numerical resolution for several values of $a$ from $1$ to $3$ was performed and the result is displayed in Figure \ref{figure:N opt}. We also present in Figure \ref{figure:var opt} a comparison between the optimal variance (with the optimal distribution $(\beta_i^*)_i$ and optimal $N$) and the classical Monte Carlo one. There we can see that for $a \lesssim 2.5$ the new estimator \eqref{eq:definition alpha} performs better in terms of variance; especially for $a < 2$ it remains finite while $\var{\mmck} = \infty$.

As explained in Section \ref{ss:Geometric randomisation} we consider now a Geometric random variable $T$ with parameter $\beta$ for the random truncation.
\begin{proposition}\label{propo:var_explicit}
If $X$ is a Pareto random variable with parameter $a > 1$ and $\forall n \in \N,\; \proba{T \geq n} = e^{-\beta n}$ then:
\[
\var{\z} = \dfrac{m (m - 1)^2}{2 \gamma(\beta,N) - m}
\]
and
\begin{equation} \label{eq:beta opt pareto parametric}
\beta_\text{opt} = \log \l \dfrac{1}{B_+} + 1 \r
\end{equation}
where $B_+$ is the positive root of the quadratic polynomial $P(B)$:
\[
P(B) = \dfrac{2N_\text{opt} - m}{(N_\text{opt}-1)^2} B^2 - 2 m B -  \l m (N_\text{opt}-1)^2 + 2N_\text{opt}^2 \r .
\]
\begin{proofprop}
One gets the expression of the variance directly from Section \ref{ss:Definition of the moment estimator} with $\gamma(N,\beta)$ instead of $N$. Then, denoting $B = 1/(e^\beta - 1)$, one solves the problem:
\[
\dfrac{\partial}{\partial B} \l \l N + B \r \l \dfrac{a}{2(a - 1) \gamma - a} \r \r = 0 .
\]\qed
\end{proofprop}
\end{proposition}
With this relation and the one of Eq. \eqref{eq:relation beta_opt N_opt} one can derive the optimal parameters $(\beta_\text{opt}, N_\text{opt})$. Figure \ref{figure:N opt} shows a numerical resolution of this problem for several values of $a \in (1, 3]$.

Furthermore, if one considers the approximation of the optimisation problem \eqref{eq:problem optim geom} with relation \eqref{eq:beta approximation} instead of \eqref{eq:relation beta_opt N_opt}, one has to minimise $N \mapsto (N^2 + N - 1) m (m - 1)^2 / (N + 1 - m)$. Denoting $N_\text{app}$ this minimiser, one has:
\begin{equation} \label{eq:N approximation}
N_\text{app} = \max \l m - 1 + \sqrt{m^2 - m - 1} , 2 \r
\end{equation}
This approximation is the red dotted-dashed line of Figure \ref{figure:N opt}. As we can see, it is in good agreement with the optimal values, both for the parameter $N$ and for the global variance (see further Section \ref{ss:Comparison of the estimators} and Figure \ref{figure:var opt}).

\subsection{Comparison of the estimators}
We have seen in Sections \ref{ss:Optimal randomisation} and \ref{ss:Geometric randomisation} two ways of implementing the \emph{ideal} estimator $\m$ defined in Section \ref{ss:Definition of the moment estimator} with a fixed computational budget. Then we have presented their exact behaviours in a case of a Pareto random variable. These two ways involve a truncation of the infinite sum \eqref{eq:mk def} by an integer-valued random variable $T$. In the first implementation the distribution of $T$ and the number $N$ of point processes are optimised in order to minimise the estimator variance. In the second implementation, the distribution of $T$ is enforced to be geometric and its parameter as well as $N$ are optimised.

While the first implementation is optimal in terms of variance, it requires to solve a combinatorial problem, which can turn it into a poorer algorithm in terms of computational time. In this scope, the parametric algorithm constraining the randomising variable $T$ to be geometric with parameter $\beta$ is much simpler to implement. The aim of this section is to benchmark these two implementations and to challenge the optimal parameters against the fixed ones we will suggest.

More precisely, while both optimisations ended up with optimal parameters depending on the distribution of $X$, we also consider the parametric algorithm with parameter $\beta_\text{app}$ given by \eqref{eq:beta approximation} and $N = N_\text{app}$, $2$, $5$ or $10$.

\begin{figure}[!ht]
\centering
\subfloat[Optimal values for $N$ in the general (cf Section \ref{ss:Optimal randomisation}) and in the parametric (cf Section \ref{ss:Geometric randomisation}) cases with the approximation of equation \eqref{eq:N approximation}.$a$ is the parameter of the Pareto distribution.]{
\hspace{9pt}
\def\svgwidth{0.45\textwidth}\input{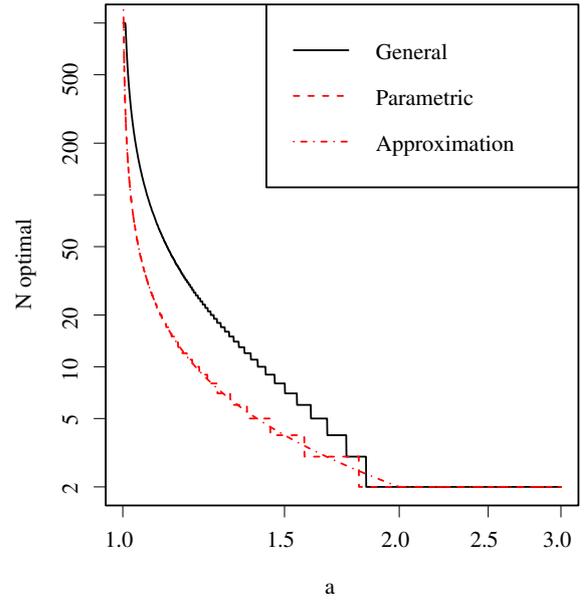} \label{figure:N opt}} \\
\subfloat[Ratios of the standard deviations of different estimators over the standard deviation of the optimal estimator $\alp$ of Section \ref{ss:Optimal randomisation}. The classical Monte Carlo estimator is defined in Eq. \eqref{eq:monte carlo definition}; $\m$ is the ideal estimator \eqref{eq:mk def}; the other estimators are randomised estimators \eqref{eq:definition alpha} with enforced geometric distribution for $T$ with parameter $\beta$ and $N$ as follows: $(\beta_\text{opt}, N_\text{opt})$: optimal parameters of Proposition \ref{propo:existence solution cas parametrique}; $(\beta_\text{app}, N_\text{app})$: approximated optimal parameters of Eq. \eqref{eq:beta approximation} and \eqref{eq:N approximation}. $a$ is the parameter of the Pareto distribution.]{\def\svgwidth{0.45\textwidth}\input{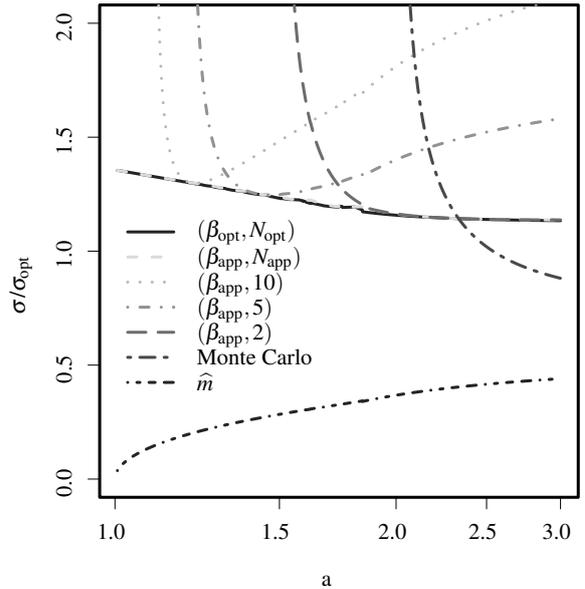} \label{figure:var opt}}
\caption{Theoretical resolution of problems \eqref{eq:optim problem} and \eqref{eq:problem optim geom} when $\proba{X > x} = 1 \wedge x^{-a}$.}
\label{figure:N et var opt}
\end{figure}

Figure \ref{figure:var opt} shows the relative increase of the standard deviations due to the suboptimal implementations for a given computational budget, $\ie$ for a given number of generated samples. It also shows the standard deviation ratios between the optimal implementation, the classical Monte Carlo estimator \eqref{eq:monte carlo definition} and $\m$ given by \eqref{eq:mk def}. For this latter, it is assumed that its computational cost is $N$, $\ie$ that it costs $1$ to simulate an increasing random walk (see Definition \ref{def:increasing rw}) while it requires an infinite number of simulated samples. This calls for certain comments:
\begin{itemize}
\item the parametric implementation with optimised parameters $(\beta_\text{opt},\, N_\text{opt})$ remains competitive against the optimal implementation (solid black line going from $\approx 1.3$ to $\approx 1.1$);
\item the parametric implementation with parameters $\beta_\text{app}$ and $N_\text{app}$ is almost not distinguishable from the parametric implementation with optimal parameters $\beta_\text{opt}$ and $N_\text{opt}$. This means that it is not necessary to strive to estimate the parameters $(\beta_\text{opt},\, N_\text{opt})$;
\item the classical Monte Carlo estimator is better than the optimal implementation as soon as $a \gtrsim 2.5$ and better than the parametric implementation as soon as $a \gtrsim 2.3$; this confirms that nested sampling is especially convenient for heavy-tailed random variables;
\item the standard deviation of $\m$ illustrates the efficiency of the ideal estimator compared to the classical Monte Carlo one (cf. Proposition \ref{propo:var_pareto}), with a standard deviation at least twice as small;
\item generally speaking and without any knowledge on the distribution of $X$, $N$ should not be set too small as the variance increases much faster when it is smaller than the optimal value; especially with $\beta = \beta_\text{app}$ finiteness condition of the variance writes $a > 1 + 1/N$.
\end{itemize}
Given these results we can consider that the parametric implementation is a good trade-off between minimal variance estimation and complexity, especially when no information on the distribution of $X$ is provided.

\section{Example}

The aim of this section is to check the consistency between theoretical formulae and practical results with non-ideal conditional sampling. It is also to demonstrate how \emph{bad} stopping criteria can alter nested sampling and how randomised estimators can resolve this issue. We first explain how we perform conditional simulation and give pseudo-code for both $\Z$ and $\alp$. Then we present results on an example from \citep[Section 18]{skilling2006nested} that we slightly modify. The presented results are obtained with $500$ simulations and boxplots extend to the extreme values.

\subsection{Simulating conditional distributions}

When no conditional sampler is available, a general idea is to use convergence properties of an ergodic Markov Chain to its unique invariant probability distribution. Assuming $\U$ is a $d$-dimensional random vector with \pdf $f_U$, it means that we intend to generate a Markov Chain with stationary \pdf $\propto \one_{g(\mathbf{u}) > x} f_U(\mathbf{u})$. This implementation is rather simple when a reversible transition kernel is available. In the sequel we make use of the transition kernel suggested by \citet{cerou2012sequential} detailed on Algorithm \ref{algo:K_def} for Gaussian input space.

\begin{algorithm}
\caption{Transition kernel for $\U \loi \mathcal{N}(0, \mathbf{I}_d)$ \citep{cerou2012sequential,guyader2011simulation}}
\label{algo:K_def}
\begin{algorithmic}
\Require initial state $\u$, $\sigma$, \burnin $b$
\While{$b > 0$}
\State Pick $\mathbf{W}$ from a standard multivariate Gaussian distribution
\State $\mathbf{U}^* \gets \dfrac{\mathbf{u} + \sigma \mathbf{W}}{\sqrt{1 + \sigma^2}}$
\If{$g(\U^*) > x$}
\State $\u \leftarrow \U^*$
\EndIf
\State $b \gets b - 1$
\EndWhile
\State \Return $\u$
\end{algorithmic}
\end{algorithm}

Because the goal is to reach the stationary state of the Markov Chain, several transitions have to be done to insure independence between the starting point and the final sample and adequacy with the targeted distribution. This number of transitions is referred to as a \textit{burn-in} parameter $b$. Eventually the last generated sample is kept. In theory, one can start from any point provided the \textit{burn-in} is large enough but practically speaking it is profitable to start with a point approximately following the targeted distribution as \textit{burn-in} will then serve mainly independence purpose. Furthermore, the step size $\sigma$ is initialised at $\sigma = 0.3$ and further updated after each use of the transition kernel -- \ie each $b$ transitions -- to get an acceptance rate close to $0.5$.

\begin{remark} \label{rem:tau avec MH}
The \textit{burn-in} parameter increases the cost of an estimator because it needs several simulations for only one sample. In this context, the computational cost defined in Proposition \ref{propo_tau} becomes $\tau = N + b T$ and is the number of calls to the generator of $X$ (which amounts to generate $\U$ and to call $g$). Since this increase is common to all algorithms considered here, we will not mention it any more.
\end{remark}

\subsection{Pseudo-code}
As explained above, we do not intend to solve the combinatorial optimisation problem in the general case and so we present here a pseudo-code for the parametric case. Reader interested in the optimal resolution is referred to \citep{rhee2013unbiased}. We then present in Algorithm \ref{algo:pseudo code zk} how to compute $\z$ and in Algorithm \ref{alpha:pseudo code alpha} how to compute $\alp(c)$. In this latter case we assume that $N$ and $\beta$ are given, being optimised (with previous knowledge or simulations) or not.
\begin{algorithm}[!ht]
\caption{Pseudo-code for $\z$}
\label{algo:pseudo code zk}
\begin{algorithmic}[2]
\Require $N$, $\beta$
\State Generate $\mathtt{T}$ according to $\proba{T \geq n} = e^{-\beta n}$
\State Generate $N$ random variables $(X_i)_{i=1..N}$ according to $\mu^X$ \label{step:generate N rv}
\State $\mathtt{times[0]} \leftarrow  0$; $\mathtt{delta[0]} \leftarrow  0$
\For{i in 1:T}
\State $\mathtt{ind} \leftarrow \argmin_j \, X_j$
\State $\mathtt{times[i]} \leftarrow X_{\mathtt{ind}}$
\State $\mathtt{delta[i-1]} \leftarrow (\mathtt{times[i]} - \mathtt{times[i-1]})\cdot \dfrac{\l 1 - 1/N \r^{i}}{e^{-\beta i}}$
\State Generate $X^* \sim \mu^X( \cdot \mid X >  X_\mathtt{ind})$
\State $X_\mathtt{ind} \leftarrow X^*$
\EndFor
\State $\mathtt{ind} \leftarrow \argmin_i \, X_i$
\State $\mathtt{times[T+1]} \leftarrow X_{\mathtt{ind}}$
\State $\mathtt{delta[T]} \leftarrow (\mathtt{times[T+1]} - \mathtt{times[T]})\cdot \dfrac{\l 1 - 1/N \r^{T}}{e^{-\beta T}}$
\State $\z = \Sum[i=0][T]\mathtt{delta[i]}$
\end{algorithmic}
\end{algorithm}

\begin{remark} \label{rem:algo z N taille pop et parametre}
Note that in Algorithm \ref{algo:pseudo code zk}, $N$ is both the theoretical parameter of the number of increasing random walks per $\z$ and the size of the population for conditional simulation purpose. Hence it should not be set too small according to the dimension of the problem. This is a side effect of this practical implementation. Alternatively one could generate several $\z_i$ sequentially to \emph{aggregate} all the samples for conditional simulations. Hence $N$ could be chosen only according to theoretical guidelines. However it would disable parallel implementation. Some recent work on the parallel implementation of Sequential Monte Carlo may be used here \citep{verge2013parallel}. Note also that it is not necessary to consider only the minimum of the $N$ samples in Algorithm \ref{algo:pseudo code zk}; however in the context of Markov Chain drawing it is better to select the starting point in a relatively big population already following the targeted distribution.
\end{remark}

\begin{algorithm}
\caption{Pseudo-code for $\alp(c)$}
\label{alpha:pseudo code alpha}
\begin{algorithmic}
\Require $c$, $N$, $\beta$
\State $G \leftarrow 0$; $\alp \leftarrow 0$;
\While{$c > 0$}
\State Generate $T^*$ according to $\proba{T \geq n} = e^{-\beta n}$
\State $c = c - (N + T^*)$; $G = G +1$; $T[G] = T^*$
\EndWhile
\If{$c<0$}
\Comment{discard the last replica if it exceeds the budget}
\State $G = G - 1$; $T = T[1:G]$
\EndIf
\For{$g$ in 1:G}
\State Start Algorithm \ref{algo:pseudo code zk} from step \ref{step:generate N rv} with $\mathtt{T} = T[g]$
\State $\alp = \alp  + \z$
\EndFor
\State $\alp = \alp / G$
\end{algorithmic}
\end{algorithm}
Basically, Algorithm \ref{alpha:pseudo code alpha} is just a wrap-up of Algorithm \ref{algo:pseudo code zk} with an update of the remaining computational budget. If one intends to use Markov Chain simulation as presented in Section \ref{ss:Simulating conditional distributions} then one has to take into account the \burnin $b$ and update $c$ in Algorithm \ref{alpha:pseudo code alpha} as follows: $c = c - (N + b T^*)$.

\subsection{Variance increase}
\begin{figure}[!ht]
\centering
\def\svgwidth{0.47\textwidth}
\input{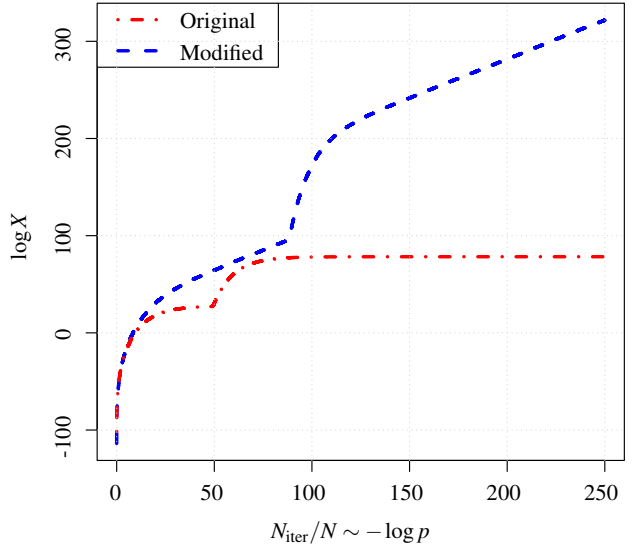}
\caption{Log-Likelihood against probability for the original example of \citet[Section 18]{skilling2006nested} (Eq. \eqref{eq:original example}) and the modified version (Eq. \eqref{eq:modified example}). Both lines are got from a sample run of nested sampling with $N = 300$ and stopping criterion $250N$ iterations.}
\label{fig:log likelihood example}
\end{figure}

In this section, we intend to check the variance increase between the ideal estimator $\m$ of Section \ref{ss:Definition of the moment estimator} and the suboptimal randomised estimator of Corollary \ref{coro:double variance Z bar}. To do so, we use an example from \citet{skilling2006nested} where it is known that $100$ iterations per particle on average are enough. We also compute (NS) the original nested sampling estimator, \ie the estimator of Eq. \ref{eq:original infinite nested sampling}. (NS) and $\m$ differ only in the weights used: $\exp -1/N$ instead of $1-1/N$ ; thus they are computed in the same run. The aim is to estimate the evidence of a likelihood with uniform prior over a $d-$dimensional unit cube: $m = \E{g(\U)} = \E{X}$ with:
\begin{equation} \label{eq:original example}
g(\u) = 100 \prod \limits_{i = 1}^d \dfrac{e^{-u_i^2 \big/ 2 u^2}}{\sqrt{2 \pi} u}
+ \prod \limits_{i = 1}^d \dfrac{e^{-u_i^2 \big/ 2 v^2}}{\sqrt{2 \pi} v} ,
\end{equation}
$\U \sim \mathcal{U} \l -[\tfrac{1}{2}, \tfrac{1}{2} ]^d \r$, $d = 20$, $u = 0.01$ and $v = 0.1$. This represents a Gaussian ``spike'' of width $0.01$ superposed on a Gaussian ``plateau'' of width $0.1$. Figure \ref{fig:log likelihood example} plots the log-likelihood $\log x$ against the log-tail distribution $\log p_x$.

\begin{figure}[!ht]
\centering
\input{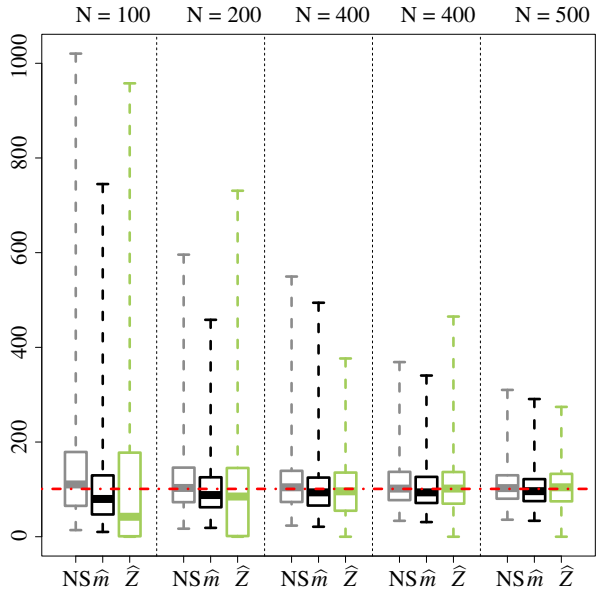}
\caption{Boxplots of ideal \emph{infinite} nested sampling $\m$ of Eq. \eqref{eq:mk def} and (NS) of Eq. \eqref{eq:mk link nested} and randomly truncated $\Z$ (Corollary \ref{coro:double variance Z bar}) for the estimation of $\E{g(\U)}$ with $g$ as in Eq. \eqref{eq:original example} and $\U \sim \mathcal{U} \l -[\tfrac{1}{2}, \tfrac{1}{2} ]^d \r$, $d = 20$. (NS) Ideal nested sampling is got with $N_\text{iter} = 100N$ as this is known to be enough in this case. (NS) and $\m$ are obtained from the same runs. The (red) dot-dashed line is the theoretical value of $m$.}
\label{fig:double var}
\end{figure}

We then run nested sampling with stopping criterion ``number of iterations = $100N$'' as well as $\Z$ for several values of $N$ from $100$ to $500$. Figure \ref{fig:double var} shows the boxplots of the estimators. On the one hand $\Z$ has good convergence properties, on the other hand the bias and variance increase due to the original nested sampling weights is clearly visible. Table \ref{table:variance increase} summarises these numerical results: both $\Z$ and $\m$ are unbiased while (NS) has a bias of order $1/N$ (cf Remark \ref{rem:variance increase original ns}). The variance increase between $\m$ and $\Z$ is in good agreement with the theoretical relationship of Corollary \ref{coro:double variance Z bar}, it is $\var{\Z}(N) = \var{\m}((N+1)/2) \approx 2 \var{\m}$. Also the ratio $\var{\mathrm{NS}}/\var{\m}$ goes from $1.14$ to $1.9$. This variance increase appears to be of order $1/N^2$, which is consistent with the variance increase between $\widehat{p_x}$ and $\widetilde{p_x}$ (see Remark \ref{rem:suboptimal p estimator}).

Hence, the optimal choice of the nested sampling weights leads to significant variance reduction and removes the bias of the original nested sampling when it goes \emph{far enough}. Unbiasedness can be maintained at the cost of at most doubling the variance of the estimator and even less compared to the currently used nested sampling weights. Furthermore, there is no need to choose (and justify) a stopping criterion for nested sampling any more.

\begin{table}[!ht]
\centering
\renewcommand{\arraystretch}{1.3}
\begin{tabular}{|c|>{$}c<{$}|>{$}c<{$}|>{$}c<{$}|>{$}c<{$}|>{$}c<{$}|}
\hline
N & 100 & 200 & 300 & 400 & 500 \\
\hline
$\E{\text{NS}}$ & 142.3 & 117.7 & 114.6 & 111.5 & 109.5 \\
$\E{\m}$ & 103.0 & 100.8 & 102.8 & 102.8 & 102.6 \\
$\E{\Z}$ & 111.9 & 97.4 & 100.4 & 103.7 & 102.4 \\
$\var{\Z}/\var{\m}$ & 3.23 & 2.49 & 1.90 & 2.20 & 1.70 \\
$\var{\text{NS}}/\var{\m}$ & 1.90 & 1.33 & 1.24 & 1.17 & 1.14 \\
$\var{\Z}/\var{\text{NS}}$ & 1.71 & 1.87 & 1.54 & 1.87 & 1.5 \\ 
\hline
\end{tabular}
\caption{Variance increase between the randomised unbiased nested sampling estimator $\Z$, the original biased nested sampling (NS) and the ideal unbiased estimator $\m$.}
\label{table:variance increase}
\end{table}

\subsection{Adaptive stopping criteria}

As we stated in the Introduction, one of the main concern of this paper was to point out the potential risk of using nested sampling with a \emph{bad} stopping criterion. In this context we run nested sampling on the previous example with the adaptive stopping criteria mentioned in Section \ref{ss:Parallel implementation}. The first one is directly picked out from \citep{chopin2010properties}, it is ``stop when the current increment is less than $10^{-8}$ times the current estimate''. The second one is based on the estimation of the information $H$ and is the one described in the Appendix of \citep{skilling2006nested}; it is ``stop when the number of iterations is greater than $2 N H$''. Figure \ref{fig:double var} shows that for $N = 500$ the estimators should be well converged and so we set $N = 500$.

\begin{figure}[!ht]
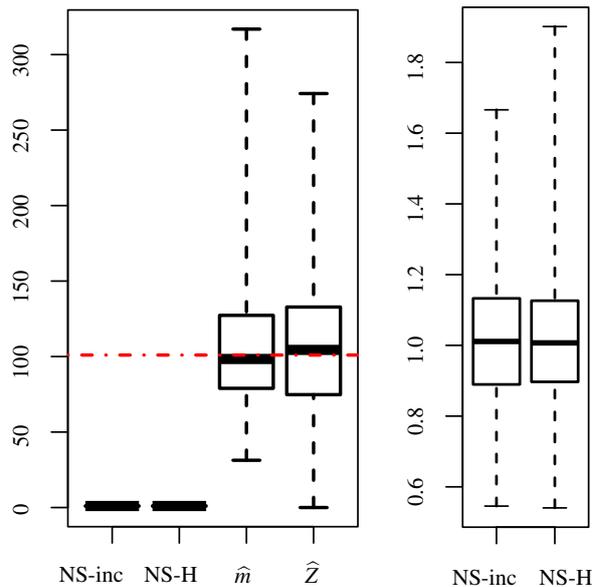

\centering
\subfloat[Nested sampling estimators with adaptive stopping criteria, $\m$ and $\Z$]{
\def\svgwidth{0.27\textwidth}
\input{CopFig5.tex}
}
\hspace{0.01\textwidth}
\subfloat[Zoom on the nested sampling estimators with adap\-tive stopping criteria]{
\def\svgwidth{0.163\textwidth}
\input{CopFig6.tex}
}
\caption{Effect of the choice of a stopping criterion for nested sampling estimator when estimating $\E{g(\U)}$ with $g$ as in Eq. \eqref{eq:original example} and $\U \sim \mathcal{U} \l -[\tfrac{1}{2}, \tfrac{1}{2} ]^d \r$, $d = 20$. (NS-inc): nested sampling stopped when current increment is less than $10^{-8}$ times the current estimator; (NS-H): nested sampling stopped when the number of iterations exceeds $2 N H$; $\m$ and $\Z$ as in Figure \ref{fig:double var}. The (red) dot-dashed line is the theoretical value of $m$.}
\label{fig:adaptive criteria}
\end{figure}

Figure \ref{fig:adaptive criteria} shows that nested sampling estimator can be not consistent if the termination rule is not well-chosen. Here both implementations miss the spike. In this context, the random truncation of $\Z$ appears as a conservative practice. However, even though $\z$ allows for parallel computing \citep[cf.][Section 4.2]{walter2015moving}, $\z$ as well as the adaptive stopping criteria do not let work with a fixed computational budget. Yet one may have to work with fixed computational resources.

\subsection{Nested sampling with fixed computational budget}

There is only one nested sampling implementation which allows for fixing the total computational budget in advance. It is the one which stops after a given number of iterations. Following \citet{rhee2013unbiased} we have proposed in Sections \ref{ss:Optimal randomisation} and \ref{ss:Geometric randomisation} a randomised estimator which also works with a predetermined computational budget. It is still unbiased and supports a Central Limit Theorem. The goal of this section is to compare these two estimators. We slightly modify the previous example \eqref{eq:original example} to narrow the spike: $u = 0.001$ instead of $u = 0.01$, and to make the random variable heavy-tailed:
\begin{equation} \label{eq:modified example}
g_\text{ht}(\u) = g(\u) / \l \Sum[i = 1][d] u_i^2 \r^{0.4 d}.
\end{equation}
Figure \ref{fig:log likelihood example} compares this modified example with the original one. The heavy-tailed behaviour with tail index $1/0.8 = 1.25$ is clearly visible (limit slope of log-likelihood is $0.8$) as well as the effect of the narrower spike (shift of the mass from $-\log p \approx 50$ to $- \log p \approx 90$). With Inv-$\chi^2$ approximation of $1/\sum U_i^2$, the sought value is $\E{g_\text{ht}(\U)} \approx 1.08 \times 10^{42}$.

Nested sampling is run with $N = 1000$ and $N = 10000$. We stop it after $100 N$ iterations as in \citep{brewer2011diffusive}. This makes a total computational budget $c = 10^5$ (resp. $10^6$). $\alp$ is implemented with a suboptimal geometric randomising variable with parameter $\beta_\text{app}$ (Eq. \eqref{eq:beta approximation}) and $N = 20$. According to Remark \ref{rem:algo z N taille pop et parametre}, $N = d$ because it is both the theoretical parameter of $\alp$ and the population size for conditional sampling. Considering the heavy-tail behaviour of $X = g(\U)$, the estimator has a finite variance as soon as $a > 1 + 1/N = 1.05$. One the one hand we know here that the tail-index of $X$ is equal to $1/0.8 = 1.25$; on the other hand it is easy to check this condition afterwards by estimating the slope on the plot $\log X$ against $N_\text{iter}/N$ as in Figure \ref{fig:log likelihood example}.

\begin{figure}[!ht]
\centering
\def\svgwidth{0.47\textwidth}
\input{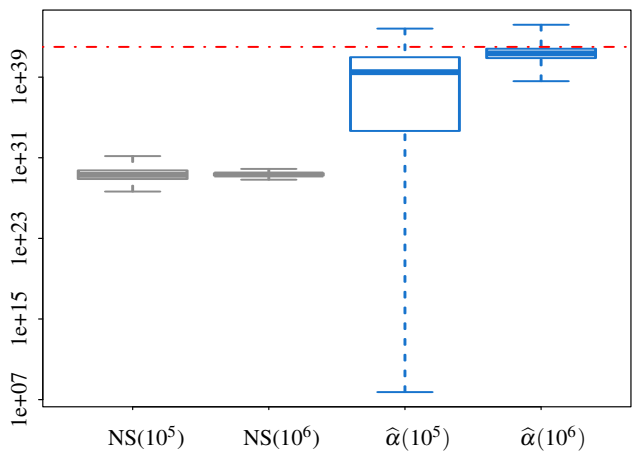}
\caption{Estimation of $\E{g_\text{ht}(\U)}$ with $\U \sim \mathcal{U} \l -[\tfrac{1}{2}, \tfrac{1}{2} ]^d \r$, $d = 20$. (NS): nested sampling stopped after $100N$ iterations; $\alp$: estimator of Section \ref{ss:Geometric randomisation} with $\beta_\text{app}$ (Eq. \eqref{eq:beta approximation}) and $N = 20$. $10^5$ and $10^6$ are the computational budgets used. The (red) dot-dashed line  is the theoretical value of $m$.}
\label{fig:NS et alpha}
\end{figure}

It is visible on Figure \ref{fig:NS et alpha} that nested sampling did not go far enough and misses an important part of the mass: $\E{\mathrm{NS}(10^5)} =  5.32 \times 10^{29}$ and $\E{\mathrm{NS}(10^6)} =  2.41 \times 10^{29}$ while the reference value is $1.08 \times 10^{42}$. On the other hand, $\alp$ is unbiased (estimated means are $6.43 \times 10^{41}$ and $1.52 \times 10^{42}$). However, it does not seem to be approximately Gaussian yet. Indeed $\Z$ can be relatively heavy-tailed \citep{mcleish2011ageneral} and a consequent computational budget may be required for $\alp$ to effectively become normally distributed.

\section{Conclusion}
Nested Sampling has been proposed as a method for estimating the evidence in a Bayesian framework and applied with success in a great variety of areas like astronomy and cosmology. Since its introduction, a lot of work has been done to clarify its convergence properties \citep[\eg][]{evans2007discussion, chopin2010properties,keeton2011statistical} and to handle the issue of conditional sampling \citep[\eg][]{mukherjee2006nested,brewer2011diffusive, martiniani2014superposition}. However nested sampling termination remains an open issue and a matter of user judgement \citep[Section 7]{skilling2006nested}.

Linking nested sampling with recent results in rare event simulation, this paper extends it to the estimation of the mean of any real-valued random variable (being bounded or not) and goes on step further by giving the optimal nested sampling weights and proving that
\begin{enumerate*}[label=\arabic*)]
\item an idealised nested sampling with slightly modified weights and an infinite number of iterations is unbiased;
\item its variance is always lower than the classical Monte Carlo estimator one's; and
\item the random variable of interest does not need to have a finite second-order moment to produce an estimator with finite variance.
\end{enumerate*}
This latter property makes nested sampling especially relevant for heavy-tailed random variables as developed Section \ref{s:Application to heavy-tailed random variables}.

Furthermore, we also present two ways of implementing a practical unbiased estimator with an a.s. finite number of terms, resolving the issue of choosing an arbitrary stopping criterion. The first estimator can be used exactly as usual nested sampling and preserves unbiasedness while only doubling the variance of the ideal estimator (infinite number of terms). The second one can be used with a predetermined fixed computational budget and supports a Central Limit Theorem. Practically speaking, they both enable parallel implementation (unlike usual adaptive nested sampling strategies) and do not depend on the random variable of interest.

As for any nested sampling implementations, they require to be able to generate samples according to conditional laws and theoretical results are derived with this hypothesis. In some cases, exact conditional sampling may be possible. When the random variable of interest is the output of a computer code, Markov Chain drawing like Metropolis-Hastings algorithm can overcome this issue. If only \iid samples are available, further work has to be done to explicit the link between the increasing random walk presented in Section \ref{ss:exteme event simulation} and, for example, Pareto-type distributions. 

\begin{acknowledgements}
The author would like to thank his advisors Josselin Garnier (University Paris Diderot) and
Gilles Defaux (Commissariat \`a l’Energie Atomique et aux Energies Alternatives) for their advices and suggestions as well as the reviewers for their very relevant comments which helped improving the manuscript. This work was partially supported by ANR project Chorus.
\end{acknowledgements}

\appenproof
\Closesolutionfile{ann}
\input{demo}

\renewcommand{\section}{\oldsection}
\bibliographystyle{spbasic}      

\bibliography{biblio}   

\end{document}